\documentclass[a4paper]{article}

\usepackage[english]{babel}
\usepackage[utf8x]{inputenc}

\usepackage[a4paper,top=3cm,bottom=2cm,left=3cm,right=3cm,marginparwidth=1.75cm]{geometry}

\usepackage{amsmath}
\usepackage{graphicx}
\usepackage[colorlinks=true, allcolors=blue]{hyperref}

\renewcommand{\k}{{\vec k}}
\newcommand{\q}{{\vec q}}
\newcommand{\p}{{\vec p}}
\newcommand{\x}{{\vec x}}
\title{Contextual Wavefunction Collapse:\\An integrated theory of  quantum measurement}
\author{Barbara Drossel\\
Institute for Condensed Matter Physics, Technische Universit\"{a}t Darmstadt\footnote{email: drossel@fkp.tu-darmstadt.de }\\ \\
George Ellis\\
Mathematics Department, University of Cape Town\footnote{email: george.ellis@uct.ac.za}}

\begin{document}
\maketitle

\begin{abstract}
\it{This paper is an in depth implementation of the proposal \cite{ellis2012} that the quantum measurement issue can be resolved by carefully looking at top-down contextual effects within realistic measurement contexts. The specific setup of the measurement apparatus determines the possible events that can take place. The interaction of local heat baths with a quantum system plays a key role in the process. In contrast to the usual attempts to explain quantum measurement by decoherence,  we argue that the heat bath follows unitary time evolution only over limited length and time scales \cite{drossel2017ten} and thus leads to localization and stochastic dynamics of quantum particles that interact with it. We show furthermore that a theory that describes all the steps from the initial arrival of the quantum particle to the final pointer deflection must use elements from classical physics. This proposal also provides a contextual answer to the puzzle of the origin of the arrow of time when quantum measurements take place: it derives from the cosmological Direction of Time. Overall, our proposal is for Contextual Wavefunction Collapse (CWC).}
\end{abstract}

\section{Introduction}\label{sec:intro}
The problem of interpreting quantum mechanics and in particular quantum measurement arises when quantum physics is considered to be a universal theory that describes all of matter, including the (macroscopic) measurement apparatus used for the quantum measurement. A way out of this dilemma consists in postulating limits to the validity of quantum mechanics \cite{leggett,ellis2012}.  However, this poses the problem of identifying where these limits lie, as quantum phenomena are not confined to the microscopic scale. 

In discussions of the issue, often the following picture of a measurement is used: By interacting with the detector, the quantum system is becoming entangled with an environment that has a macroscopic number of degrees of freedom. By tracing over these degrees of freedom, the reduced density matrix of the quantum system becomes that of a mixed state. When the Hamiltonian that describes the interaction with the environment is chosen appropriately, this mixed state is characterized by the probabilities of the different possible outcomes of the measurement process. 

However, as emphasized by various authors, this type of consideration leaves many open questions. In particular, it does not explain why in a one-time experiment a single, definite, outcome is obtained \cite{adler2003decoherence,schlosshauer2005decoherence}.
Furthermore, a measurement, as performed in the real world, is more complex than described above. It involves several steps, from initial quantum transitions within the detector to an amplification process and finally a macroscopic change in a read-off unit. The steps of this process depend on specific local physical conditions that set up the context for the measurement process. 

In this paper, we therefore want to present an integrated approach that focuses on the context dependence of the measurement process \cite{ellis2012}. Our basic view 
is that measurement is truly a stochastic, nonunitary process that can be labeled as a "projection to an eigenstate" of an observable $A$ \cite{Dirac1982,isham2001lectures,greenstein2006quantum} - a "wave function collapse".  The intrinsic heat bath of the detector plays a key role in this process, as it is specified by a limited number of bits (as implied by the entropy formula) and therefore follows unitary time evolution only over limited length and time scales  \cite{drossel2017ten}. Beyond these scales, the system behaves classically.  We will show this view is viable by considering the different steps of a measurement in sufficient detail and by analyzing the physical models and equations that are used for the  description of each step. We will show that apart from the very first stage, none of these steps can be described purely by unitary quantum mechanics, but requires recourse to elements from classical physics. Even though the context of the quantum system can be described quantum mechanically at least for some of the degrees of freedom and up to certain length and time scales, the wider environment is a classical environment.  

\subsection{The measurement problem} \label{sec:intro_measure}
There is a key tension in quantum physics between the unitary evolution described by the Schr\"{o}dinger equation, and the measurement process usually described as a projection operation (summaries and references are given in \cite{isham2001lectures,greenstein2006quantum,ellis2012}; for the density matrix version, see (3.68)-(3.71) in \cite{witten2018mini}).\\

\textbf{ Unitary Evolution} 
When the quantum system to be measured and the measurement apparatus both are described quantum mechanically, the Schr\"{o}dinger equation  prescribes the time evolution of the wave function $|\Phi(t)\rangle$ of this combined system. This time evolution has the form 
\begin{equation}  \label{eq:unitary1}
 \textbf{U}:\,|\Phi(t_1)\rangle = U(t_1,t_0) |\Phi(t_0)\rangle \, ,
 \end{equation}  
 where $U$ is invertible.
Since the Schr\"odinger equation is linear in the wave function, the time evolution gives rise to macroscopic superpositions of quantum objects. 

Let us describe the wave function of the quantum system as
\begin{equation}
|\Psi\rangle = \Sigma_ic_i |\psi_i\rangle\, , \label{wavefunction}
\end{equation}
where $\{|\psi_i\rangle\}$ is a complete basis associated with an (observable) operator $A$, and let us  
describe the measurement apparatus including all its internal (environmental) degrees of freedom by a many-particle wave function  $|\mathcal{E}\rangle$. 
If the measurement apparatus is set up such that it measures the observable $A$, and if the quantum system is prepared in an eigenstate of $A$, the time evolution of an initial state $|\Phi(t_0)\rangle \equiv |\psi_i\rangle |\mathcal{E}_0\rangle$ at time $t_0$ leads to a final state $|\Phi(t_1)\rangle =|\psi_i\rangle |\mathcal{E}_i\rangle$ at time $t_1$, with the apparatus state $|\mathcal{E}_i\rangle$ including a macroscopic pointer that indicates that the quantum system has been measured in the state $i$. 

However, when the initial state of the quantum system is a superposition of the form (\ref{wavefunction}) with more than one of the $c_i$ being nonzero, the outcome of the time evolution according to the Schr\"odinger equation must be
\begin{equation}
U(t_1,t_0)|\Psi(t_0)\rangle |\mathcal{E}_0\rangle = \Sigma_ic_i U(t_1,t_0) |\psi_i\rangle |\mathcal{E}_0\rangle = \Sigma_ic_i |\psi_i\rangle |\mathcal{E}_i\rangle \, , \label{unitary}
\end{equation}
which is a superposition of macroscopically different states. \\

\textbf{Measurement} In contrast, a measurement experiment of the observable $A$ gives a definite outcome, which is a single eigenstate $|\psi_n\rangle$ of $A$ for some $n$. The specific eigenstate $n$ and corresponding eigenvalue $\lambda_n$  is not determined by the dynamics, but the probability for outcome $n$ is given by $|c_n|^2$ (the Born rule). This projection 
 \begin{equation}
 \Pi:\,\,|\Phi(t_0)\rangle= \Sigma_ic_i |\psi_i\rangle |\mathcal{E}_0\rangle \rightarrow |\Phi(t_1)\rangle = |\psi_n\rangle |\mathcal{E}_n\rangle \label{collapse}
 \end{equation} 
 is an irreversible process as all the other terms from equation (\ref{unitary}) are lost, hence we cannot reconstruct the wave function $|\Psi(t_0)\rangle$ prior to measurement from $|\Phi(t_1)\rangle$ afterwards (equivalently, $\Pi$ is not an invertible map). The measurement process thus cannot be described by unitary time evolution according to  the Schr\"{o}dinger equation. However the wave function has no meaning unless such measurement events take place, so quantum theory needs some account of this process. A key point  is that wave function collapse (\ref{collapse}) takes place in many contexts that are not in fact `measurements', i.e. laboratory situations leading to a pointer outcome. We shall therefore, after analysing in detail in  Sections 2 to 5   the processes whereby actual measurements happen, consider in Section \ref{sec:discussion_comments}   the much more general contexts where `events' (\ref{collapse})  occur.\\
 
\textbf{Proposed solutions} There are essentially two classes of proposed solutions to this dilemma: 
The first class (Section \ref{sec:first}) are those that adhere to linear superposition and unitary time evolution even for macroscopic systems consisting of $10^{23}$ particles and much more (even the entire universe). In this case, the relation between calculations of the type (\ref{unitary}) and our observation of a single, stochastic measurement outcome (\ref{collapse}) must be re-interpreted.  The second class (Section \ref{sec:second}) are those that postulate limits to unitary time evolution and linear superposition. In this case, the challenge consists in identifying where the limits of validity lie. The key underlying issue is discussed in Section \ref{sec:intro_issue}. How this paper tackles the issue is outlined in Section \ref{sec:intro_thispaper}.

\subsection{The first class of solutions}\label{sec:first}

The first class of explanations of the measurement process, which consider the mathematical description according to the Schr\"odinger equation to be the most appropriate and comprehensive description of the time evolution of a nonrelativistic system, includes 
the many worlds \cite{everett1957relative},  relational \cite{rovelli1996relational}, consistent (or decoherent) histories \cite{griffiths1984consistent}, modal, de Broglie-Bohm, and statistical \cite{Ballentine} interpretations. These interpretations differ in the ontological status that they ascribe to the wave function and in their explanation of the observed randomness. Thus, for instance, the proponents of the consistent (or decoherent) histories interpretation \cite{griffiths1984consistent,griffiths2017quantum,gell1993classical,hartle2011quasiclassical} do  not consider the wave function as real but consistent, stochastic histories, which are calculated from the standard formalism by using projection operators and decoherence functionals. 
Everett \cite{everett1957relative} considers the wave function as real but claims that our consciousness takes a stochastic route through the various branchings that occur at each measurement event. The  relational interpretation \cite{rovelli1996relational,rovelli2018space} argues that the actual values of the physical variables of a system are only meaningful in relation to another system with which this system interacts.
Similarly, modal interpretations are not concerned with an objective description independent of observers, but only in consistent classical descriptions of systems that are on a microscopic level described by quantum mechanics \cite{hollowood2017decoherence}. The statistical interpretation \cite{Ballentine} suggests that wave functions do not describe single systems but ensembles of identically prepared systems. The de Broglie-Bohm pilot wave theory is different in nature as it postulates hidden variables, namely the particle position and its deterministic time evolution, which depends non-locally on the wave function. The stochasticity of all measurement events of the future time evolution of a particle is encoded in the initial value of the hidden variable.   

An important role in most of these interpretations is played by decoherence theory \cite{zurek2003decoherence}, which provides good reasons to claim that the environmental states $\{\mathcal{E}_i\}$ that result from the combined time evolution (\ref{unitary}) of system and apparatus are orthogonal to each other, such that the  reduced density matrix for the measured system alone becomes diagonal and thus represents a classical ensemble of measurement outcomes. However, critics argue that these calculations, even when supplied with the various interpretations, cannot pin down the transition from a (potentially classical) superposition to a definite, random outcome of an individual measurement \cite{adler2003decoherence,schlosshauer2005decoherence,ellis2014evolving}. But if science shall explain what happens in nature, it must explain why there are specific outcomes. Quantum events are not always averaged when going to macroscopic systems (such as in thermodynamics), but individual events can change the time evolution of macroscopic systems, for instance when a mutation induced by cosmic radiation causes cancer and later death in a person.  
Even the mathematical calculations of decoherence theory are being challenged on the grounds that the calculations always include assumptions such as statistical independence and "typicalness" of the degrees of freedom of the environment that are foreign to a deterministic theory \cite{kastner2014einselection,drossel2015relation} and must be added to it as an additional assumption. In other words, the environment is assumed to be in a random state among all possible ones, and this introduces stochasticity into the system \cite{gisin2017collapse}.

\subsection{The second class of solutions}\label{sec:second}

The second approach to the contradiction between the quantum and classical world consists in accepting limits of validity to the description of many-particle systems by the Schr\"odinger equation \cite{petrosky1993poincare,leggett1992nature,ellis2012,bassi2013,drossel2017ten,gisin2016}. In fact, as emphasized by T. Leggett \cite{leggett1992nature} and R. Laughlin \cite{laughlin2000theory}, quantum mechanical calculations in condensed matter physics are never done using the full Hamiltonian, but by using effective theories. Furthermore, they depend on classical concepts and classical environments for those degrees of freedom that are not explicitly modeled \cite{ellis2012,chibbaro2014reductionism}. The view that there are limits to a unitary time evolution leads to interpretations such as the traditional Copenhagen interpretation and various collapse theories \cite{penrose1996gravity,bassi2013}. 

The Copenhagen interpretation makes a fundamental distinction between the classical, macroscopic  world and the quantum world. It has the strength that it is empirically adequate and that it takes into account top-down effects from the macroscopic, classical world on the quantum world, such as occur in all the experiments in quantum optics laboratories and in quantum computing where individual qubits are set by the experimenter. The Copenhagen interpretation has, however, the problem that it does not identify the boundary between the classical and quantum worlds.

Stochastic collapse models introduce a nonlinear, stochastic term that has significant effects only when a macroscopic number of particles are involved. They have the problem that they are ad-hoc. They postulate a more "fundamental", more "microscopic" reality underneath the description by unitary quantum mechanics. They do not give a formalism depending on the nature of local physical macro systems, such as a measuring apparatus, as one might expect. 

\subsection{The underlying issue}\label{sec:intro_issue}

A satisfactory and empirically adequate answer to the measurement problem, in our view, can only be found when taking seriously the insight that every physical theory has a limited range of validity, which is true also for quantum mechanics, as it cannot do without classical physics \cite{chibbaro2014reductionism,drossel16quantumclassical} (see 
Section \ref{sec:problems:classical_ingredients}  below). We therefore address the question on what scale or for which type of systems quantum theory holds, i.e.~a theory described by a  wave function $\psi$ that evolves unitarily according to the Schr\"{o}dinger equation (\ref{eq:unitary1}).

It is commonplace nowadays to take it for granted that it is meaningful to talk of the wave functions of macroscopic objects, including cats and even the entire universe. However while there has undoubtedly been extraordinary progress in demonstrating quantum effects over macroscopic scales, such as entanglement of photons over distances of the order of 100's of kilometers \cite{entangledZeil2007}, these can only occur under most exceptional circumstances, when the quantum system is sufficiently isolated from the rest of the world. On our view  there is no meaningful wave function of the universe, or even for a cat,  in the former case because there is no context in which it attains any useful meaning, in the latter case {\it inter alia} because of non-linearities in its physiological relationships at all levels \cite{ellis2012}, and because the internal degrees of freedom and the surrounding environment, acting (among other effects) as a heat bath, ensures it behaves as a classical system \cite{drossel2017ten} (see Section \ref{sec:problems-heat_bath} below). 

Related criticism is brought forward by Walter Kohn, who points out that  a wave function of $10^{23}$ particles is not a legitimate scientific concept, since it can neither be prepared nor measured with sufficient accuracy \cite{kohn1999}.

\subsection{This paper}\label{sec:intro_thispaper}

The view taken here is close to the Copenhagen interpretation: it assumes firstly that the wave function is real (i.e., descriptive and objective rather than epistemic)
and secondly that the classical world is as real as the quantum world, and has a top-down effect on the quantum world \cite{ellis2012}\cite{EllisTopDown}. We do not add any \textit{ad hoc} terms into the Schr\"{o}dinger equation to attain our results.  We show rather that wave function collapse can be obtained using the formalism developed by the theory of open quantum systems when one abandons the untestable and implausible claim that the environmental heat bath can be described by an infinite-precision wave function that is subject to unitary time evolution. Instead, we interpret the time evolution of the reduced density matrix as that of a quantum system with a limited temporal and spatial coherence, which means that the time evolution of the reduced density matrix must be unravelled in terms of a wave function of the quantum system. This leads to the trapping of the wave function in one of the blocks of the density matrix, since the density matrix goes to a blockdiagonal form with time (Section \ref{HB_unravel}). The trapping of the wave function in one of the blocks is the most important step towards a definite measurement outcome. Furthermore, all this implies that the quantum system is not entangled with the environmental degrees of freedom beyond the scale of quantum coherence. This makes the detector classical on larger length scales and thus explains 
quantum contextuality: the possible measurement outcomes are determined by the measurement apparatus. 
\\

In order to describe the interplay between the classical and quantum worlds in detail, we will use a concrete model of quantum measurements with four ingredients: the system (particle) to be measured; a macroscopic structure that constitutes the measurement apparatus; a metastable setup of the apparatus; and the heat bath and  heat sink coupled to the apparatus (internal and external). The particle is described quantum mechanically; the apparatus is described classically; the heat bath is responsible for making the apparatus classical on length scales beyond the thermal wavelength and for localizing its atoms and thus preventing linear superpositions of different measurement results, and the heat sink takes up energy produced during the process. Measurement events are possible because the world is not in equilibrium, as a result of the macroscopic context. The local structure extracts for example  electrons and moves them elsewhere, and hence makes the process irreversible. 
The ultimate source of the arrow of time in this process is traced to the cosmological arrow of time. \\

In outline, Section \ref{sec:building_blocks} looks at the building blocks of measurement. Section \ref{sec:model-steps} looks at mathematical modeling of the different steps. Section  \ref{sec:two-slit} looks at the 2-slit experiment in the light of this approach, and shows how Born's rule arises. Section  \ref{sec:problems} looks at key features  in the physics of the process,  including the origin of the arrow of time. Section \ref{sec:discussion} summarizes the proposal, and considers further applications where this approach could be useful, considering `events' as well as quantum measurements. Overall we emphasize the contextual nature of the processes taking place.

\section{The building blocks of 
measurement} \label{sec:building_blocks}

There is a great contrast between the abstract way measurements are presented in standard texts on quantum theory such as Dirac \cite{Dirac1982}, Feynman and Hibbs \cite{FeynmanandHibbs}, and Isham \cite{isham2001lectures}, and real measurement contexts. We believe that the route to a full understanding of the measurement problem, bridging this gap, is to consider measurement in the case of specific real detection situations (Section \ref{sec:building_blocks_context}) and the common features in those cases (Section \ref{sec:building-blocks-common}), and then to generalise to general principles that are likely to hold in all realistic cases (Section \ref{sec:building-blocks-elements}).

\subsection{Examples of real measurement contexts} \label{sec:building_blocks_context}

As mentioned above, collapse to an eigenstate, as represented by Equation (\ref{collapse}), is often called a ``measurement''. However in fact such collapses happen far more frequently than during laboratory measurements. We will refer to such happenings as ``events''. 
Real measurements with some laboratory apparatus, as we discuss now,  are a subset of "events". We will return to a discussion of ``events'' as defined here in Section \ref{sec:discussion_comments}.\\

Here are a variety of examples in order to illustrate that real measurements involve common features:

\begin{itemize}
\item \textit{Photographic plate}: impact of a particle releases an electron (or several electrons) that moves to the location of a defect in the silver bromide crystal and attracts a silver ion, producing neutral silver. These neutral silver atoms are later made visible by the development procedure.

\item \textit{Photodiode}: electrons are lifted from the valence band to the conduction band and are extracted via the applied voltage. If the photodiode is operated in avalanche mode, this electron will trigger an avalanche of electrons that can be measured as a macroscopic current.

\item \textit{Photo multiplier}: electrons are accelerated repeatedly from target plates via high voltage differences. This is indeed a cascade of detection events, in each case the freed electrons being removed by the electric field to go on to the next plate.

\item \textit{Geiger counter}: a gas becomes ionized by an incoming particle. The applied voltage causes an avalanche of subsequent ionizations that results in a noticeable current signal in the detector, similarly to the photomultiplier. 

\item \textit{Cloud chambers and bubble chambers}: here, atoms or molecules that are  ionized by incoming particles trigger the condensation of droplets. This is again an amplification process. The liquid in a bubble chamber is prepared by being put in a superheated, metastable phase.

\item \textit{Rhodopsin Molecule in an eye}: freed electrons cause a complex chain of changes in rhodopsin 
{as in phototransduction}
that eventually lead to an action potential that travels along the optic nerve to the brain
{from sensory neurons}. 
Note that if the molecule is isolated rather than being in its biological context, it does not function as a photodetector.
\item A \textit{Charge Coupled Device} (CCD), such as is used to make images in all cellphones, collects electrons in bins associated with each pixel in a 2-dimensional array, and then reads them out by electrically transferring them along a readout line to a charge amplifier, which converts the charge into a voltage. The sequence of voltages thus created is then sent to a digital or analogue signal processing device. 
\item  \textit{Spin measurements}, as in the Stern-Gerlach experiment and in quantum computing. 
In the former case, essentially the position of the atom (upper or lower trajectory) is measured by a photographic plate (or a set of two detectors). 
\end{itemize}

\subsection{The common features} \label{sec:building-blocks-common}

There are a series of common features across all these cases:
\begin{enumerate}
\item The apparatus must have a macroscopic structure (i.e. arrangement of atoms) that is sufficiently stable before the measurement. Some component of this structure makes a reliable transition to a new state that is again stable enough that the measurement result can be read off. The system must thus initially be in a metastable state, making a transition to a more stable state in response to an incoming particle or photon. 

\item An electron or ion is removed from the interaction location because of the context in which the detection event takes place, for instance by extracting it via a potential gradient. 

\item An amplification process, for instance an avalanche, makes the event visible on a macroscopic scale and thus allows the readout of the measurement result. For most of the above examples, this amplification follows immediately after the previous step. For the photographic plate and the CCD, the amplification step is decoupled from the first step and can occur much later, during development of the photographic plate or during the readout process.

\item Since the apparatus is macroscopic and at finite temperature, it includes a heat bath with a macroscopic number of degrees of freedom. Decoherence theory deals with the question of how this heat bath makes the reduced density matrix of the measured system classical  but does not solve the puzzle of how unique measurement outcomes arise. We will therefore need to discuss the nature of the heat bath further below. 

\item The process described in steps 2 and 3 generates some heat that is given to the environment. Since the process is irreversible, the entropy of the universe is increased. Therefore the relevant environment includes not only the internal heat bath of the apparatus, but also an external heat sink.   

\item In each case there is a reset process whereby the detector is readied for the next detection, usually by replacing the electrons that were released in the detection process. It is during this reset process that the Landauer cost $k T \ln 2$ per bit of the detection, or more, is extracted. This process requires energy from the environment. This reset step also confirms the irrevocable irreversibility of the process as a whole, as any remaining information about the previous state is over-written. In the case of the photographic plate, the reset process is more expensive as it requires to introduce an entire new plate.
\end{enumerate}

All this means that measurement is a top-down effect from the apparatus  to the relevant particles \cite{ellis2012},
since it is the setup of the apparatus that determines the type of transitions they can undergo and the possible measurement results.   
It also depends on the environment of the apparatus, which allows irreversible processes to take place by acting as a heat sink.

\medskip

In the following, we will choose one specific example, the photodiode, in order to make a theoretical description of the measurement process that includes all these steps. We will write down concrete models for these different steps (Section \ref{sec:model-steps}). Then we will generalize this approach to the 2-slit experiment (Section \ref{sec:two-slit}).
Note that no single model can do the job, and that the models required typically mix classical and quantum elements.

\subsection{Required elements for modeling the measurement process }\label{sec:building-blocks-elements}

 We use a simplified picture of the detector (Section \ref{sec:detector_model}) and measurement process (Section \ref{sec:detector_processes}) that includes all the relevant elements described above. This picture is inspired by the process of photodetection by a photodiode operated in avalanche mode. Later, we will extend this picture to many detectors such that the double slit experiment can be described. We do not invent a new theory, but we build on available theories and methods for describing the different steps. In contrast to many other treatments of the measurement problem, we emphasize the need to combine different theories, and we pin down the places where the transition from quantum to classical behavior occurs.  We show that apart from (part of) the first step nothing is described by using unitary time evolution of a wave function. 

\subsubsection{Detector model}\label{sec:detector_model} We model the detector by the following features
\begin{itemize}
\item  Bound states  and a conduction band for electrons, resulting from the macroscopic structure of the detector. The events that happen when a photon hits the detector are thus due to a top-down effect enabled by the nature of the detection apparatus \cite{ellis2012}.
When the photon is absorbed, an electron is lifted from a bound state to the conduction band, which is an energy continuum. More electrons can be lifted to the conduction band by collision with the first electron during the avalanche process.
\item An internal heat bath which interacts with the electron that is in the conduction band and localizes it. This heat bath is made up by the phonons of the semiconductor material, again existing due to its structure, and represents the environment in the context of decoherence theory. How this heat bath localizes the electron and limits the scope of unitary time evolution will be discussed in detail.
\end{itemize}
\begin{itemize}
\item  Furthermore, we must assume at least implicitly an  external heat sink, which makes the entire process truly irreversible and represents the shared environment for the system of several detectors used later for modeling the double slit experiment.
\item An electrical field that removes the electron from the site of the excitation. Since this conduction of the electron occurs in the semiconductor material, we need again a coupling to the phonon environment, which is responsible for the electrical resistance. 
\end{itemize}
\subsubsection{Detector processes}\label{sec:detector_processes}
The steps of the measurement process, each of which requires mathematical modeling, are then the following:
\begin{enumerate}
\item The electron is lifted from a bound state into the continuum (e.g. the conduction band) by an incoming electromagnetic wave. In order to model this step, a suitable basis of quantum states is chosen. 
\item  The electron becomes and remains localized in the conduction band. In order to describe this localization, a Lindblad-type equation is required, which depends on an external environment. As we will discuss below, this is the crucial step where the event becomes definite, and it requires ingredients beyond standard decoherence arguments. We will explain two of these ingredients: On the one hand, we will argue that a heat bath cannot be fully described quantum mechanically and that therefore standard decoherence considerations are inappropriate. Instead, the heat bath should be viewed as consisting of local wave packets with limited quantum coherence, and this leads to stochasticity.  On the other hand, we will argue that a feedback between the electron and the environment will lead to a nonunitary time evolution of the combined system+environment.
\item The electron moves through the semiconductor by a standard electrical conduction process. This requires a look at the theory of Ohm's law. We shall explain that no theory of Ohm's law is fully quantum, but always semiclassical.
\item The electron interacts with other electrons and triggers thus a macroscopic avalanche. Again, no pure quantum description of such a process is available. The best theories involve semiclassical Boltzmann equations. 
\item The final detection step, which is the (classical) detection process of the macroscopic current, is done using classical physics (for instance classical electrodynamics and classical mechanics if in the end a mechanical pointer is moved.)
\item The reset process needs the external heat sink and requires energy. Because it is essentially equivalent to state vector preparation, it is a non-unitary process \cite{ellis2012}. 
\end{enumerate}
The next section develops this model in some detail. 

\section{Mathematical modeling of the different steps}\label{sec:model-steps}

Now, let us collect the theories that are available to describe all these listed steps. We discuss the steps in the same order and with the same numbering as in the previous section, in subsections 3.1 to 3.6. 
We  emphasize that all the theories have elements of classical physics, and all are context dependent.
\subsection{Photon-triggered state transition}
We focus at first on single-photon detection and assume therefore that one photon enters the diode. We consider one electron that is originally in the bound state and can be lifted to the conduction band due to the interaction with the photon (the existence of the conduction band is of course due to the environment of the specific crystalline structure that is the context of the measurement). 
This process can be modeled by using two types of states, namely a state with one photon with wave vector $\vec k$ and polarization $\lambda$ and a ground-state electron and a state with zero photons and an excited electron with wave vector $\p$. The  transition that we consider is thus
\begin{equation}
|1_{\vec k},g\rangle \to |0,e_\p\rangle \label{1g0e}
\end{equation}
By using an obvious notation for the creation and annihilation operators of the various states, the Hamiltonian for the time evolution reads in second quantization 
\begin{equation}
H= H_{ph}+H_{el} + H_{int}=\hbar \omega_{\vec k} a^\dagger_{\vec k,\lambda} a_{\vec k,\lambda} +\epsilon_g b_g^\dagger b_g + \sum_{\vec p}\epsilon_p b^\dagger_{\vec p} b_{\vec p} + \sum_{\vec p}g_{\vec k \lambda,\vec p} a_{\vec k\lambda}b_{\vec p} ^\dagger b_g + h.c. \label{H}
\end{equation}
Usually, one proceeds from here by directly calculating the transition matrix element and from there a transition rate. In order to obtain correct matrix elements, the coupling $g_{\vec k \lambda,\vec p}$ must be expressed in terms of dipole moment of the initial state and the amplitude of the photon at the location of the electron. However, our focus is different, because we are interested in the question  of when an "event" becomes definite and  not in a specific number for the transition probability. We will only need to know that  $g_{\vec k \lambda,\vec p}$ is proportional to the amplitude of the photon at the location of the interaction. We will not proceed by using Fermi's Golden rule or some higher-order calculation in order to obtain a transition probability.  In any case, the calculation of such transition probabilities involves a projection on a final state, thus circumventing our central question of how the "collapse" implied in this projection occurs.

Instead, we consider the generic form of the quantum mechanical time evolution of the system and make the ansatz
\begin{equation}
\psi(t) = \alpha(t)e^{-i(\omega_\k+\epsilon_g/\hbar) t} |1_{\vec k},g\rangle+ \sum_{\vec p} e^{-i\epsilon_p  t/\hbar}\beta_{\vec p}(t) |0,e_{\vec p}\rangle \label{superposition}
\end{equation}
with coefficients $\alpha, \beta_\p$ and with the initial condition $\alpha(0)=1$ and $\beta_\p(0)=0$. In the absence of an electron-photon interaction, the coefficients $\alpha$ and $\beta$ do not change in time. Starting from  the Schr\"odinger equation $i\hbar \frac \partial {\partial t}\psi(t) = H\psi(t) $ with $H$ given by (\ref{H}), one obtains the following coupled equations for the time evolution of the coefficients $\alpha$ and $\beta$,
\begin{eqnarray}
i\hbar \dot \alpha &=& \sum_\p g_{\vec k \lambda,\vec p}^* \beta_\p \; e^{i\left(\omega_\k-\frac{\epsilon_e-\epsilon_g}{\hbar}\right)t}\nonumber\\
i\hbar \dot \beta_\p &=& g_{\vec k \lambda,\vec p} \alpha \, e^{-i\left(\omega_\k-\frac{\epsilon_e-\epsilon_g}{\hbar}\right)t}\, .\label{alphabeta}
\end{eqnarray}
By multiplying the first line with $\alpha^*$ and the second line with $\beta_\p^*$ and considering also the complex conjugate of the resulting equations, one can easily show that 
$$
\frac d{dt} \left(|\alpha|^2 + \sum_\p |\beta_\p|^2 \right)= 0\, ,
$$
as it must be. 
The photon interacts with the detector for a limited time $\tau$, leaving the system in a superposition  (\ref{superposition}) with both terms being nonzero. 

Now, the usual procedure in calculations of quantum transitions, chosen for instance when deriving Fermi's Golden rule, would consist in considering an interaction time $\tau$ that is so short that $\alpha$ does not change, and inserting the solution of the second line of (\ref{alphabeta}) into the first one. Performing the $t$  integration then gives a delta function that ensures energy conservation $\epsilon_e=\epsilon_g + \hbar \omega_\k$ for each possible transition.  

In our case, the situation is more complicated as the photon cannot be absorbed without involving a third partner in the transition. Without this third partner, it is impossible to simultaneously achieve energy and momentum conservation. In our measurement device, this third partner is the crystal lattice of the semiconductor, which can take up the surplus momentum via the  motion of ions, i.e., via the phonon heat bath. (The energy uptake of the bath can be neglected due to the much larger mass of ions compared to electrons.) In order to keep the calculations focused on the most essential steps, we will in the following simplify equations (\ref{alphabeta}) by implementing energy conservation directly and dropping the time-dependent factors.  Furthermore, we will take into account the fact that the incoming photon is not a plane wave but a wave packet, and similarly the excited electron will not be a plane wave but a wave packet, i.e., a superposition of states with different $\p$. We will therefore replace the sum $\sum_\p g_{\vec k \lambda,\vec p}^* \beta_\p$ by an effective term $\beta g^*$, where $\beta$ is interpreted as the amplitude of the electron wave packet and where we assume that $g_{\vec k \lambda,\vec p} $ does not change in a significant way over the relevant interval of $\p$ values. The simplified equations then are 
\begin{eqnarray}
i\hbar \dot \alpha &=& g^* \beta \nonumber\\
i\hbar \dot \beta &=& g \alpha \, .\label{alphabetasimple}
\end{eqnarray}

The interaction with the phonon bath will be considered in the next subsection. This interaction does not only help to satisfy momentum conservation, but more importantly helps to transform the superposition (\ref{superposition})  to a definite outcome. This definite outcome is the detection of the photon with probability $|\beta(\tau)|^2 = 1-|\alpha(\tau)|^2$ and non-detection with probability $|\alpha(\tau)|^2$.  (Note: since there is more than one electron that can make the transition, we will need to expand our discussion. This will be done further below. Additionally, we will expand our discussion to many detectors and thus to position measurements in  section \ref{sec:two-slit}.)

\subsection{Electron localisation} \label{subsec:localization}

Once the electron is in the conduction band, it is no longer in a bound, stationary state, and becomes subject to interactions with other degrees of freedom of the semiconductor material, in particular with phonons (which exist because of the crystal context). We therefore need to describe next how the interaction with the phonons (a) definitely fixes the electron in the bound or unbound state and (b) causes an ongoing localization of the electron while being in the conduction band. A wave packet that describes an unbound electron becomes broader and broader if it is allowed to evolve freely without interactions. However, electrons in finite-temperature, Ohmian conductors are usually modeled using a semiclassical description based on wave packets (see for instance the discussion in \cite{solyom2007fundamentals2} Ch.~21).  

We will proceed in three steps in order to obtain a suitable equation for the density matrix of the electron that achieves both above mentioned goals (a) and (b). First, we will assume that the electron is in the conduction band and describe ongoing localization (Section \ref{sec:3_localization}). Second, we will extend the density matrix to include also the bound electron state in order to describe additionally that the transition (or non-transition) becomes definite (Section \ref{sec:3_definite}).  Third, we will generalize to many electrons that interact with the incoming photon, each of which could become excited to the conduction band (but at most one of them) (Section \ref{sec:3_generalisation}). 
The resulting equations will be interpreted in Section \ref{sec:problems-heat_bath} and in particular in part \ref{HB_unravel}, where we will argue that they describe indeed a collapse and not a quantum system that is entangled with the environment.

\subsubsection{Localization of one electron that is in the conduction band}\label{sec:3_localization}
The effect of the phonon degrees of freedom on the electron can be generally captured by writing a nonunitary equation for the time evolution of the density matrix of a single electron. Usually, such nonunitary equations of the Lindblad type are derived or are argued to be derivable by starting from a unitary time evolution of the considered system (here: the electron) and the environment (here: the many phononic degrees of freedom), and by taking the trace over the environmental degrees of freedom, see for instance \cite{breuer2002theory}. However, such equations can also be derived by imposing the minimal logical requirements, which are symmetries, relevant degrees of freedom, positive-definiteness and conservation of probability of the density matrix, and the semi-group property of the time evolution operator \cite{weiss2012quantum} (this is the theory of Kraus operators). In this case, no specific model of the environment is required, and even not the assumption that the environment is a quantum system.
This will become important further below when we need to discuss the interpretation of the outcome of the time evolution of the density matrix and the nature of the phonon heat bath. 

The equation available in the literature to describe the ongoing localization of the electron is an extension of the Caldeira-Leggett \cite{caldeira1983path} equation, which is for instance given in \cite{breuer2002theory} and \cite{bassi2013}:
\begin{equation}\label{eq:Lindblad}
\dot \rho = -\frac i \hbar [H,\rho] -\frac{i\gamma} \hbar [\x,\{\p,\rho\}] - \frac{2m\gamma k_BT}{\hbar^2} [\x[\x,\rho]] -\frac{\gamma}{8mk_BT}[\p,[\p,\rho]] 
\end{equation}
where $\rho$ is the density matrix of the electron and $T$ is the temperature of the environment. The Hamilton operator $H$ contains the kinetic energy and possibly further contributions to unitary time evolution due to the interaction with the environment, and in absence of the other terms it would lead to a continuous increase of the width of the electron wave packet. The last three terms can be summarized to give the expression $$\gamma\sum_{j=1}^3\left(A_j\rho A_j^\dagger- \frac 1 2A_j^\dagger A_j \rho - \frac 1 2\rho A_j^\dagger A_j \right) = \sum_{j=1}^3\gamma\left(A_j\rho A_j^\dagger- \frac 1 2\left\{A_j^\dagger A_j, \rho\right\}\right)$$
with $$A_j= \frac {x_j }\lambda + \frac {i\lambda} \hbar p_j = \frac {x_j} \lambda + \lambda \partial_j$$ and $$\lambda = \frac \hbar{\sqrt {4mk_BT}}$$ and a rate constant $\gamma$ that depends on the strength of coupling between the electron and the heat bath. Note the similarity of $A_j$ with the annihilation operator of the one-dimensional harmonic oscillator! Just as for the annihilation operator of the harmonic oscillator, the eigenfunctions of $A_j$ are Gaussian wave packets,
\begin{equation}
A_j e^{-(\x-\x^0)^2/2\lambda^2+i\k\cdot \x} = \left(\frac{x_j^0}\lambda +ik_j\lambda\right) e^{-(\x-\x^0)^2/2\lambda^2+i\k\cdot\x}
\end{equation}
The width of these wave packets is given by $\lambda$, which is half the thermal de Broglie wave length of an electron at temperature $T$. In the absence of $H$, the  stationary solutions of (\ref{eq:Lindblad}) are the eigenfunctions of $A_j$, i.e., Gaussian wave packets. Equation  (\ref{eq:Lindblad}) thus describes the combined effect of two competing processes: the delocalization of the electron wave packet by the action of $H$, and  the localization by the action of $A_j$. The crystal structure is anisotropic, i.e. introduces preferred directions. But this is not relevant for our purpose, because the thermal wavelength of the electron is much larger than the lattice spacing.

Note that an arrow of time has been introduced here already, as Eqn.~(\ref{eq:Lindblad}) is not time reversible. 

\subsubsection{Making the transition definite}\label{sec:3_definite}
So far, we have not included the bound electron state in the time evolution following the interaction with the photon. In the following, we denote the state $\psi(\tau)$ of equation (\ref{superposition}) in a simplified manner as $\alpha |g\rangle + \beta |e\rangle$, dropping the photon from the notation of the ground state since it has moved on elsewhere and is not relevant anymore. The electron density matrix at the beginning of the interaction with the phonons is therefore given by 
\begin{equation}
\rho = |\alpha|^2|g\rangle \langle g| + |\beta|^2|e\rangle \langle e| + \alpha\beta^*|g\rangle \langle e|+ \alpha^*\beta|e\rangle \langle g|  \equiv \rho_g + \rho_e + \rho_{int}   \, ,
\end{equation}
with the last term being the quantum mechanical interference term that makes the density matrix non-classical.
Interaction of the phonons with the ground state of the electrons is responsible for the dark count rate, which we neglect here, assuming that it is small. We therefore model the ground state as not interacting with the phonons, i.e., as an eigenfunction of $H$, and we set $[H,\rho_g]=0$ and $A_j|g\rangle = A_j^\dagger |g\rangle=0$.
By taking the bound state into account, the time evolution (\ref{eq:Lindblad}) of the density matrix thus becomes generalized to 
\begin{equation}\label{eq:Lindblad2}
\dot \rho = \dot \rho_e +\dot \rho_{int} = -\frac i \hbar [H,\rho_e] +\gamma\sum_j\left(A_j\rho_e A_j^\dagger- \frac 1 2\left\{A_j^\dagger A_j, \rho_e+\rho_{int}\right\}\right) .
\end{equation}
The time evolution of the interference term is therefore
\begin{equation}
\dot \rho_{int} = - \frac\gamma 2 \sum_j\{A_j^\dagger A_j, \rho_{int}\}= - \frac\gamma 2 \sum_j\{\alpha^*\beta A_j^\dagger A_j|e\rangle \langle g| + h.c.\}\, ,
\end{equation}
which describes a steady decrease of the interference term, since $A_j^\dagger A_j$ is a positive semidefinite matrix and $|e\rangle$ is never exactly an eigenstate of $A_j$ to the eigenvalue 0. 
After  a sufficiently long time, it has gone to zero, and we are left with a density matrix 
\begin{equation}
\rho = \rho_g + \rho_e = |\alpha|^2|g\rangle\langle g|+ |\beta|^2|e(t)\rangle\langle e(t)|\, , \label{rhoend}
\end{equation}
  the time evolution of which is given by $\rho_g=const$ and 
\begin{equation}\label{eq:Lindblad2a}
\dot \rho_e =  -\frac i \hbar [H,\rho_e] +\gamma\sum_j\left(A_j\rho_e A_j^\dagger- \frac 1 2\left\{A_j^\dagger A_j, \rho_e\right\}\right) \, .
\end{equation}
This represents a classical combination of a bound electron, which occurs with probability $|\alpha|^2$, and an electron in the conduction band that occurs with probability $| \beta|^2=1-|\alpha|^2$ and undergoes ongoing localization. We postpone the interpretation of this result to section  \ref{sec:problems-collapse}. There, we will argue that this describes indeed a``collapse'' of the electron wave function to a definite result. We will argue that the heat bath cannot be described by pure quantum mechanics but by a collection of localized wave packets that show quantum coherence only over limited time and length scales, and therefore the electron wave packet becomes also localized, avoiding the interpretational problems arising in the context of decoherence theory. In this way, our considerations apply to a single electron detection event, even though it is a density matrix calculation.
  Again an arrow of time occurs here.  Its source has to do with the nature of the heat bath in its cosmological context, see Section \ref{sec:arrow}. 

\subsubsection{Generalization to many electrons}\label{sec:3_generalisation}
When  there are $N$ electrons, all of which can make a transition to the conduction band due to the incoming photon, the time evolution (\ref{superposition}) becomes 
\begin{equation}
\psi(t) = \alpha(t) e^{-i(\omega_\k+\epsilon_g/\hbar) t}|1_{\vec k},g_1,\dots,g_N\rangle+ \sum_{i=1}^N  \beta_{i}(t)e^{-\epsilon_e t/\hbar} |0,g_1,..,g_{i-1}e_{i},g_{i+1},\dots,g_N\rangle \, \label{superposition2}
\end{equation}
where we have again summarized the different momentum contributions to the excited state of each electron into one term in order to simplify notation. The second and third term of the Hamiltonian (\ref{H}) are now replaced with sums over all electrons. 
The time evolution of the coefficients $\alpha$ and $\beta_i$ becomes
\begin{eqnarray}
i\hbar \dot \alpha &=& \sum_i g_i^* \beta_i \nonumber\\
i\hbar \dot \beta_i &=& g_i \alpha \, .\label{alphabeta2}
\end{eqnarray}
Here, we have again implemented energy conservation and simplified the notation for the coupling coefficients $g$, as we have done before when going to equations (\ref{alphabetasimple}). 
Assuming that all electrons are equivalent, we can make all $g_i$ identical, $g_i=g$, and all $\beta_i$ identical apart from a phase factor due to the different positions of the electrons, $\beta_i=\beta e^{i\k\x_i}$. By defining $\tilde g = \sqrt N g$ and $\tilde b = \sqrt N b\langle e^{i\k\x_i}\rangle$, we can then write (\ref{alphabeta2}) in the form
\begin{eqnarray}
i\hbar \dot \alpha &=& \tilde g^* \tilde\beta \nonumber\\
i\hbar \dot {\tilde\beta} &=& \tilde g \alpha \, .\label{alphabeta2mod}
\end{eqnarray}
The set of $N$ electrons thus behaves like a single electron, but with a larger coupling coefficient and therefore a faster increase of the amplitude of the excited electron  state.

So far, the time evolution of the system is reversible, and  the coherent superposition in the second term of (\ref{superposition2}) can in principle return to the ground state by re-emitting the photon. Such processes have indeed been observed \cite{gisin2017experimental} for superpositions of more than $10^{10}$ electrons in specifically designed experimental setups where the interaction with the phonon bath is sufficiently weak. In the case of a detector, we have the opposite situation where the interaction with the phonon bath rapidly closes the time window for coherent re-emission.  We therefore proceed again by calculating the consequences of the interaction of the system (\ref{superposition2}) with the phonon bath. The time evolution of the density matrix contains now a sum over all electrons on the right-hand side of (\ref{eq:Lindblad2}), and the interference term now includes interference between the excited states of different electrons. If we assume that the effect of the heat bath on the different electrons is uncorrelated between the electrons, the time evolution now goes to a blockdiagonal form where all interference terms between different electrons have vanished, in addition to the interference between the ground state and the excited state of each electron. This means that the density matrix assumes the form
\begin{equation}
\rho = \rho_g + \sum_{i=1}^N \rho_{ei} = |\alpha|^2|g_1,\dots, g_N\rangle\langle g_1,\dots,g_N|+ \sum_{i=1}^N|\beta_i|^2|e_i(t)\rangle\langle e_i(t)|\, , \label{rhoend2}
\end{equation}
which means that we obtain a long-term behavior described by an equation (\ref{eq:Lindblad2a}) for each electron, with the different equations being decoupled from each other. 
The probability that electron number $i$ becomes excited is given by $|\beta_i|^2$. Again, the interpretation of this result and the explanation how this describes wave function collapse is referred to sections  \ref{sec:problems-collapse} and \ref{sec:problems-heat_bath}. In section \ref{sec:two-slit}, we will generalize the results further in order to cover the double-slit experiment. Then we will see that a calculation of this type gives the Born rule for position measurements.

\subsection{Electron conduction} \label{sec:electron_conduction}

Conduction of the electron is described on the most microscopic level by linear response theory. Generally, linear response theory describes the response of an observable (here: the current density) to an applied external field (here: the electrical field) that couples to a system variable (here: the charge density). This response is expressed in terms of a susceptibility (here: the electrical conductivity).
In a general notation:  If $A$ is the operator the field couples to (charge density) and $B$ the operator corresponding to the measured observable, the dynamical susceptibility that describes the response is given by \cite{schwabl2005advanced}
\begin{equation}
\chi_{AB}(t-t') = \frac i \hbar \theta(t-t') \langle[\hat A(t),\hat B(t')]\rangle_0\, .\label{chiAB}
\end{equation}
The expectation value is taken with respect to the quantum state of the system and using an appropriate thermodynamic equilibrium ensemble (if $T>0$). 

Interestingly, linear response theory breaks explicitly time reversal invariance and invokes the concept of causality \cite{frisch2014causal}, an issue which we will take up again in section \ref{sec:problems-causality}.
The situation is even more complicated  than this: In order to evaluate (\ref{chiAB}) for the case of electron conduction, one does not employ a straightforward quantum mechanical calculation of expectation values and partition functions based on a purely microscopic description.  In the words of Tony Leggett \cite{leggett1992nature}, 
\begin{quote}\it No-one has ever
come even remotely within reach of deriving Ohm's law from microscopic
principles without a whole host of auxiliary assumptions ('physical
approximations'), which one almost certainly would not have thought of
making unless one knew in advance the result one wanted to get.\end{quote} The best known and simplest derivation of Ohm's law is obtained from the Drude model, in which electrons are considered as classical particles that are repeatedly accelerated by the electrical field and lose velocity due to  scattering events, which are accounted for only implicitly by introducing a relaxation time. The most detailed type of derivation of Ohm's law is for instance given in \cite{solyom2007fundamentals2}, ch. 24 by describing the electron as a wave packet and using a Boltzmann equation for describing collisions with phonons (see section 24.2, pages 368f and consider the case of constant temperature and chemical potential). 

 When we denote the (space- and time-dependent) electron density with $f$, the Boltzmann equation for electrons in an electric field has the form 
\begin{equation}
\frac{\partial f}{\partial t}+\vec v_{\vec k}\cdot\frac{\partial f}{\partial \vec r}-\frac{\partial f}{\partial\epsilon_{\vec k}}\vec v_{\vec k}\cdot e\vec E=\left(\frac{\partial f}{\partial t}\right)_{coll} \label{BE}
\end{equation}
with the collision term being due to collisions with lattice defects and phonons and  the left-hand side describing the total time derivative.  If we ignore lattice defects (for instance by assuming a perfect crystal), the collision term depends only on phonons and has the form
\begin{eqnarray}
\left(\frac{\partial f}{\partial t}\right)_{coll} &=& \frac 1  V\sum_{\vec k',\vec q,\lambda}\bigl\{W_{\k';\k,\q,\lambda}f(\k')[1-f(\k)][1+g_\lambda(\q)] \nonumber\\&+& W_{\k',-\q,\lambda;\k}f(\k')g_\lambda(-\q)[1-f(\k)]\nonumber\\
&-& W_{\k;\k',\q,\lambda}f(\k)[1-f(\k')][1+g_\lambda(-\q)] \nonumber\\ &-&W_{\k,-\q,\lambda;\k'}f(\k)g_\lambda(-\q)[1-f(\k')]
\bigr\}\nonumber 
\end{eqnarray}
with the various scattering cross sections $W$ and 
with a phonon distribution function $g$ according to thermal equilibrium. 
This equation is based on electron density and not on single electrons. But since the electrons don't interact with each other in this theory, the Boltzmann equation can also be used for a single electron by interpreting it as an equation for the probability density. 

Interestingly, the author mentions that more generally the phonon distribution function is affected by collisions with the electrons. This means that there is a feedback between electrons and phonons, which will be discussed further below in Section \ref{HB_feedback}.

The Boltzmann equation is based on a Markov assumption, since no memory effects are taken into account. This implies that those degrees of freedom that are not considered explicitly can be supposed to be in equilibrium, which means that they have forgotten the past.   Furthermore, equation (\ref{BE}) breaks time reversal symmetry. 
Below, in Sections \ref{sec:problems-causality} and \ref{sec:arrow}, we will discuss these interesting issues further.

\subsection{Amplification}\label{sec:amplification} 
The amplification process makes the transition triggered by the photon macroscopic. In the specific context considered by us, this amplification process takes place in form of an avalanche, where electrons in the conduction band induce transitions of further electrons to the conduction band. The general type of model suitable to describe such a process uses a position-dependent rate equation for the density of bound and excited electrons. Parameters in this equation are the mean electron velocity $v$ and a transition rate $\alpha$ that depends on the cross section for the excitation of electrons by other excited electrons, 
\begin{eqnarray}
\frac{\partial n_e(z,t)}{\partial t} &=& -v \frac{\partial n_e(z,t)}{\partial z} + \alpha n_e(z,t)(n_b(z,t))\nonumber \\
\frac{\partial n_b(z,t)}{\partial t} &=& -\alpha n_e(z,t)(n_b(z,t))\, .\label{avalanche}
\end{eqnarray}
Again, a Markov assumption is made and time reversal symmetry is broken. 
\subsection{Detector output} 
\label{sec:detector_output}
A macroscopic pointer of some kind indicates that a current is flowing, which means that a photon has been detected. The pointer deflection can be obtained by an analog, classical mechanism. For instance, the current that results from the avalanche process can be  sent to a moving coil ammeter. In this device, the current goes through a (macroscopic) coil with $N$ turns and face area $A$ that can rotate in a magnetic field $\vec B$, which we assume to be homogeneous in order to keep equations simple.  The torque on the coil due to the current $I$ is $NI\vec A\times \vec B$, with the direction of the vector $\vec A$ pointing along the coil axis (i.e., it is perpendicular to the face area). The torque causes the deviation of the pointer from its equilibrium orientation by an angle $\theta$, which is counteracted by springs. Assuming a linear restoring force $-c\theta$ and a friction force $-\eta \dot \theta$, the equation of motion of the angle $\theta$ (to which the pointer is attached so that $\theta$ can be read off) is 
\begin{equation}
J\ddot \theta = NIA B\cos \theta -\eta \dot \theta - c\theta\, ,
\end{equation}
with $J$ being the moment of inertia of the coil. This is a classical equation framed in terms of classical concepts. \\

In digital devices, the current leads for instance to emission of light by an LED (which can for this purpose be described also in  classical language).

\subsection{Reset}\label{sec:reset}

The reset process, which restores the initial metastable state mentioned in Section \ref{sec:building-blocks-common}, is essentially a process of state preparation and hence is non-unitary and so cannot be described by the Schr\"{o}dinger equation \cite{ellis2012}. It is therefore achieved by some operational aspect of the macroscopic apparatus. The way it works is context dependent, but in general is of the nature of filling again all vacated sites with electrons to achieve the desired uniform initial state, with minimum information represented (a `blank slate'). 

Landauer's principle \cite{landauer1961irreversibility}  holds that any logically irreversible manipulation of information, such as the erasure of a bit, must be accompanied by a corresponding entropy increase in non-information-bearing degrees of freedom of the information-processing apparatus or its environment \cite{bennett2003notes}. Thus the reset process will cost an energy of at least $\Delta E = k T \ln 2$ per bit.

The way this works out will be context dependent: the energy  required for the reset process will need to be investigated in each case by relating it to the number of bits involved in the reset process, which overwrites whatever was there in the relevant locations and hence is a process of deleting bits. 

\section{Two slit experiment}\label{sec:two-slit}
Let us now generalize the modeling done in Section \ref{sec:model-steps} to a 2-slit experiment. We consider the situation that only one photon  at a time is sent through the double slit. Usually, a photographic plate or a screen is used in this experiment in order to make the interference pattern visible that results after a sufficient number of photons have been sent through the slits. In order to apply the considerations of the previous section, we replace the screen by a set of many identical photodiodes, which together cover the entire area of interest. Each of the detectors has its location $\vec r$ in the two-dimensional plane and covers a small area $a$. We will show that at each event, each detector will click with a probability proportional to the intensity of the interference pattern at the location of the detector. 

Let us denote the photon amplitude at the location $\vec x_n$ of detector number $n$ with $A(\vec x_n)$.  The coupling coefficients $g_{in}$ of the electrons in detector $n$ to the photon are then proportional to  $A(\vec x_n)$. 
Equation (\ref{superposition2}) now contains a sum over $n$ in addition to the sum over $i$, and the state vectors contain the states of all electrons in all detectors,
\begin{equation}
\psi(t) = \alpha(t) e^{-i(\omega_\k+\epsilon_g/\hbar) t}|1_{\vec k},g_{11},\dots,g_{MN}\rangle+ \sum_{i=1}^N  \sum_{n=1}^M\beta_{ni}(t)e^{-\epsilon_e t/\hbar} |0,g_{11},..,e_{ni},\dots,g_{MN}\rangle \, .\label{superposition3}
\end{equation}
If we assume that all detectors are identical and all electrons within a detector are equivalent, we can again drop the index $i$ in the coupling coefficients $g$ and the amplitudes $\beta$. For many detectors, equations (\ref{alphabeta2mod}) take then the form (again with $\tilde \beta_n = \sqrt N \beta_n\langle e^{i\k\x_{ni}}\rangle_i$ and $\tilde g_n = \sqrt N g_n$)
\begin{eqnarray}
i\hbar \dot \alpha &=& \sum_n\tilde g_n^* \tilde\beta_n \nonumber\\
i\hbar \dot {\tilde\beta}_n &=& \tilde g_n \alpha \, .\label{alphabeta3mod}
\end{eqnarray}
Integrating the second equation, we obtain
\begin{equation}
\tilde\beta_n(t) = \tilde g_n \int_0^t d t'\alpha(t')\, .\label{integral}
\end{equation}
Now it is important to note that the integral is the same for all detectors, since $\alpha$ is the amplitude of the state in which none of the electrons in none of the detectors is excited. This means that the amplitude $\tilde b_n$ that an electron in detector $n$ is excited is proportional to $\tilde g_n$, which in turn is proportional to the amplitude of the photon at the location of that detector, $A(\vec x_n)$. 

We can continue the calculation exactly as in section \ref{subsec:localization}. If we assume that the effect of the heat bath on the different electrons in the different detectors is uncorrelated, the joint density matrix for all electrons evolves again to a blockdiagonal form 
\begin{equation}
\rho = \rho_g + \sum_{n,i} \rho_{eni} = |\alpha|^2|g_{11},\dots, g_{MN}\rangle\langle g_{11},\dots,g_{MN}|+ \sum_{n,i}|\beta_n|^2|e_{ni}(t)\rangle\langle e_{ni}(t)|\, , \label{rhoend3}
\end{equation}
where we have denoted the total number of detectors with $M$. By deriving equation (\ref{integral}), we have shown that $\beta_n(t)$ is proportional to $A(\vec x_n)$. This means that the probability that detector number $n$ clicks (which is given by  $N\beta_n^2  = \tilde \beta_n^2$) is proportional to $|A(\vec x_n)|^2 $.  This is Born's rule for position measurements. Again, the deeper discussion of how a density matrix that contains classical probabilities is related to outcomes of one-time measurements is referred to section  \ref{sec:problems-collapse}. 

\section{Key issues in the derivation
}\label{sec:problems}
A series of key issues underlie what we have done. These are, 
where the collapse happens (Section \ref{sec:problems-collapse}); 
 the important roles of the heat bath and heat sink (Section \ref{sec:problems-heat_bath}); issues to do with causality (Section \ref{sec:problems-causality}); the importance of classical ingredients (Section \ref{sec:problems:classical_ingredients}); the significance of top-down effects from the wider context (Section  \ref{sec:top_down_effects});  and the arrow of time issue (Section \ref{sec:arrow}).

\subsection{Where the collapse happens}\label{sec:problems-collapse}
The key step is the localization process that makes an end to the linear superposition that is created by the interaction of the incoming photon with the electrons of the detectors. We have modeled this step three times, first for one electron in a superposition of the ground state and the excited state (in the conduction band), then for many electrons in a  superposition where at most one electron is in the excited state, and then for many detectors. The initial linear superpositions (\ref{superposition}), (\ref{superposition2}), and (\ref{superposition3}), which were created by the interaction with the incoming photon, interacted with their respective surrounding heat baths, leading to a reduced density matrix for the electrons that has the blockdiagonal form (\ref{rhoend}), (\ref{rhoend2}), and (\ref{rhoend3}) for the three cases. The probabilities for the different possible outcomes are given by the traces over the respective blocks and agree with Born's rule. These outcomes are either that the photon is still present and none of the electrons is in the conduction band, or the photon has been absorbed and one of the electrons in one of the detectors is in the conduction band. 

So far, our calculation follows the standard procedure in the theory of open quantum systems. The important task now consists in relating the mixed density matrices obtained by the calculations to one-time events. The density matrices describe an ensemble of events as they contain simultaneously all possible outcomes. The time evolution leading from the initial linear superposition to the final mixed state was described by a deterministic equation (\ref{eq:Lindblad2}) for the case of one electron, and by the appropriate generalized equations for more electrons. Although we ascribed this time evolution to the interaction with the heat bath, we did not yet specify the properties of the heat bath. If the heat bath was described by a many-particle wave function that evolves deterministically in interaction with the electrons, the reduced density matrix of the electrons would describe a system that is entangled with the environment, and the measurement problem would not be resolved. However, if the heat bath is not described by a deterministic, unitary time evolution, but has a stochastic time evolution and limited quantum coherence, then the deterministic time evolution of the reduced density matrix must be interpreted as describing an ensemble of electron wave functions, each if which follows a time evolution that matches one of the possible time evolutions of the heat bath. 

Our central task in the next subsection is therefore a careful discussion of the properties of the heat bath. We will argue that the heat bath cannot be described by a global wave function, and that therefore an  entangled state of electron with the entire heat bath cannot exist. Instead, the time evolution of the reduced density matrix of the electron must be interpreted by unraveling it in terms of a stochastic equation for the electron wave function. This leads to definitive, stochastic measurement outcomes, such as in Eq.~(\ref{collapse}). 

The timescale of the collapse process is given by the decoherence time. However, with our view of the heat bath this decoherence time cannot be interpreted in the usual way. Usually, decoherence time is defined as the time scale beyond which the different environmental states that occur in the entangled wave function become orthogonal. 
Here, it is the time scale over which the memory of the initial state of the heat bath is lost.

The subsequent steps described in Sections \ref{sec:electron_conduction} to \ref{sec:detector_output}  use semiclassical or classical equations. 
The analysis thus presupposes that during electron conduction and amplification the wave packets of electrons remain localized through ongoing interaction with the phonon bath, again emphasizing that the localization of the electron by the heat bath is the crucial step that makes the transition from quantum to (semi-)classical behavior.

\subsection{The important roles of the heat bath and heat sink}\label{sec:problems-heat_bath}

Heat baths play a key role in decoherence theory and in the theory of open quantum systems. In those theories, the considered quantum system is coupled to an environment that includes many degrees of freedom. When decoherence is considered in the context of the measurement process, the environmental degrees of freedom are taken to be the internal degrees of freedom of the (macroscopic) measurement apparatus. In the standard case, which we consider here, the environmental degrees of freedom are in thermal equilibrium, so the environment is a heat bath. In decoherence theory, heat baths are understood to be quantum-mechanical many-particle systems that follow a unitary time evolution. By taking the trace over those many degrees of freedom and making a couple of plausible approximations, decoherence theory arrives at a (nonunitary) equation for the time evolution of the reduced density matrix of the quantum system. This reduced density matrix is that of a mixed state. We have given an explicit expression for the reduced density matrix relevant for the situation considered by us in Eqs.~(\ref{rhoend3}) and (\ref{eq:Lindblad2a}), building on the achievements obtained by decoherence theory. However, in contrast to the decoherence community, we do not believe that the heat bath can be described by a many-particle wave function that follows unitary time evolution. Apart from the fact that such a monstrum can neither be prepared nor be measured, assuming its existence leads to the main problem related to the measurement process, namely how a single measurement event gives a definite measurement outcome \cite{adler2003decoherence}. The consistent histories approach \cite{griffiths1984consistent,gell1993classical} avoids this problem by considering stochastic families of histories and not wave functions as real, however, we do not consider this as a satisfactory way out, as will be discussed below in section \ref{sec:different}.

By ascribing to the heat bath only those properties that are known with high confidence, we will show in the following that the interpretational problem of the measurement process can be avoided. These properties are those that feature in the formalism of statistical physics, among them a limited number of distinguishable states (with the logarithm of this number given essentially by entropy), and stochastic transitions. We will argue that this means that the heat bath can be described by a collection of local wave packets that have a certain degree of quantum coherence, but their time evolution deviates from unitary dynamics beyond certain length and time scales.

In the following, we discuss in detail this nature of a heat bath (Section \ref{HB-Nature}), including the thermal wavelength (\ref{eq:thermal_lenght}) and thermal time (\ref{eq:thermal_time}); local wave packets (Section \ref{sec:local}); heat baths vs decoherence (Section \ref{HB_decohere}); heat baths and stochastic wave function dynamics (Section \ref{HB_unravel}); 
feedback  between a quantum system and heat bath (Section (\ref{HB_feedback}); and the heat sink (Section  \ref{HB_heatsink}). Heat sinks are part of the larger environment of quantum systems, beyond the internal heat bath of the measurement device. 
Heat sinks are usually ignored in decoherence theory, but we will argue that they play an important role at making quantum events definite, in particular by breaking time-reversal symmetry.

\subsubsection{The nature of a heat bath}\label{HB-Nature}

Let us first define what we mean by a heat bath. Often, quantum systems are called "thermalized" when their time evolution has led to a state in which classical observables (sum variables, correlation functions) take values as in the equilibrium ensembles of statistical physics \cite{eisert2015quantum}. Such states can already occur in systems that consist of only a few particles and in zero-temperature systems. 

In contrast, we confine the definition of a heat bath to an equilibrated system that consists of a macroscopic number of interacting degrees of freedom such that the system shows a nonzero temperature in the sense understood by thermodynamics, namely that the temperature can be measured with a thermometer, and that the system emits thermal radiation (so there must be a heat sink to receive that radiation).  \\
  
\textbf{The requirements} that must be satisfied by a system 
in order to qualify as a heat bath in this (thermodynamic) sense are \cite{drossel16quantumclassical}:

(i) the boundary of the system is irrelevant for the thermal fluctuations inside (i.e., thermalized systems are extensive with respect to their thermodynamic properties). 

(ii) The density of states is so large and transitions between different states are so numerous that the width of the energy levels (which is related to the inverse of their lifetime) is larger than their distance. This leads to random energy exchange between the internal degrees of freedom and to the heat bath being uncontrollable from outside.

We consider these properties as sufficient for an environment that induces a wave function collapse. While extensivity and the full range of other thermodynamic properties might not be necessary, the quasi-continuous spectrum and the resulting uncontrollability of the microscopic state from outside and the irreversibility due to the continuing emission of photons are necessary in our view. 
Even a microscopic true detection device should therefore be coupled to a macroscopic number (i.e., a continuum) of degrees of freedom that exchange energy with each other and with the measured quantum system. These degrees of freedom must be considered as part of the detector. The adsorption of a photon by a macroscopic object is therefore not  
in itself a sufficient criterion for a wave function collapse.\\

\textbf{Stochasticity in the heat bath}: 
The internal state transitions, coupled to the emission of photons from the surface of the heat bath, leads to stochasticity in the system. In fact, quantum statistical mechanics, which is our best theory for thermalized systems, is based on probabilities as fundamental ingredients for the calculation of partition functions. Of course, it has been suggested that these probabilities can be calculated starting from a unitarily evolving many-particle wave function \cite{eisert2015quantum}, however, this approach burdens statistical mechanics with exactly the same interpretational problems as the measurement process, and it has been criticized for instance in \cite{drossel2017ten}. Furthermore, due to the emission of photons a fully quantum mechanical description of the heat bath by unitary time evolution would need to include an ever increasing entanglement with the external world. Claiming that such a unitary time evolution occurs nevertheless has no basis in physics as an empirical science. The wave function of the heat bath plus environment can neither be controlled nor measured \cite{kohn1999}, not even in principle. Thermodynamic systems can only be controlled top-down by (classical!) macrovariables, and the second law of thermodynamics teaches us that their intrinsic dynamics merely adjusts to the values of these macrovariables, without preferring one of its internal microstates over the other. This is just the feature of multiple realisability that occurs in  top down causation \cite{auletta2008top,EllisTopDown}.\\

\textbf{Thermal wavelength}:
The stochastic dynamics leads to a limited memory of the past and a limited temporal and spatial range over which quantum coherence and linear superposition hold. Concordantly, the statistical mechanical calculations for ideal gases, for instance, result in a characteristic length scale, called the \textit{thermal wavelength}, which is for massless particles (such as phonons or photons) given by 
\begin{equation}\label{eq:thermal_lenght}
\lambda_{th}= \frac{\pi^{2/3}\hbar c}{k_BT},
\end{equation} 
with $c$ being the velocity of the considered (quasi-)particle. For nonrelativistic  particles of mass $m$ it is given by
\begin{equation}\label{eq:thermal_lenght2}
\lambda_{th}= \frac h{\sqrt{2\pi m k_BT}}\, .
\end{equation} 
These length scales feature prominently e.g. in the textbook by Kittel \cite{kittel1998thermal}. 
Beyond the thermal wavelength, the system is described by classical statistical mechanics, but when relevant scales (such as the distance between particles) become of the order of the thermal wavelength quantum effects become relevant. This fact is taught in particular in the context of Bose-Einstein condensation, where the condition that the average particle distance becomes of the order of the thermal de Broglie wave length gives (up to a factor of order 1) the correct condensation temperature.\\

\textbf{Thermal time}: Closely associated with the thermal wavelength is the \textit{thermal time} 
\begin{equation}\label{eq:thermal_time}
t_{th}=\frac{\hbar}{k_BT},
\end{equation} 
which is the time after which a thermal wave packet has moved a distance of the order $\lambda_{th}$. 

The thermal time and length thus describe the temporal and spatial range over which quantum coherence occurs. They play exactly this role also in calculations in decoherence theory, when the reduced dynamics of the quantum system is considered. (See also Sec.~\ref{sec:3_localization}, where we have used such an equation.)

\subsubsection{Local wave packets}\label{sec:local}
Since quantum correlations occur in thermalized systems only on the scales given by the thermal length and time, 
the description by wave functions can only be used on these scales. In fact, many models and calculations are actually based on local wave packets, such as the semiclassical theories referred to in Section \ref{sec:electron_conduction}, or wave packet molecular dynamics simulations or even {\it ab initio} molecular dynamics simulations, which take quantum coherence into account only on short scales. 
Furthermore, these theories assume that wave packets remain wave packets during their time evolution. This means that these wave packets become localized again and again by a "collapse" process that picks one of the different possible time evolutions of the wave packet. 

When applied to the heat bath, this means that a proper quantum mechanical view of a heat bath is a collection of localized wave packets describing phonons. Similarly, as done above, the electron in the conduction band, which exchanges energy with the phonons, is described by a wave packet that remains localized. Unitary time evolution can thus be applied only locally and over a limited time. Beyond, deviations from unitary time evolution must be taken into account. 

\subsubsection{Heat baths vs decoherence}\label{HB_decohere}

Our discussion of the heat bath shows that those properties of a heat bath that can be verified empirically are very different from the linear, unitary description that is presumed in decoherence theory. A unitary time evolution of the heat bath has no empirical basis: the state of a heat bath cannot be microscopically controlled. It can be controlled only via the macro-variables. Concordantly, statistical physics tells us that the state of a heat bath cannot be specified exactly (which would require an infinite number of bits \cite{gisin2016}), but only with the number of bits given by its entropy, which in turn depends only on a few macroscopic properties (such as temperature and density of states). 
Thus there is no way, even in principle, to prepare identical copies of a heat bath or to measure its wave function. 

Interestingly, decoherence theory must always make assumptions about the randomness or uncorrelatedness of the degrees of freedom of the bath in order to "derive" the desired mixed density matrix. 
Such assumptions are {\it ad hoc} and presuppose what they want to show, namely that quantum  correlations need to be considered only on short length  and time scales \cite{kastner2014einselection}. In contrast, when stochasticity is included as a feature of the time evolution of a heat bath phases become randomized and a memory over a limited time emerges naturally, and so do Lindblad-type equations of evolution.

Summarizing our discussion of the heat bath and the difference to decoherence theory so far, there are several fundamental problems associated with the attempt to describe the heat bath by a unitary time evolution: first, it is not even in principle verifiable experimentally; second, it is responsible for the measurement problem as it always leads to entangled states that contain all possible future time evolutions of the quantum system that interacts with the bath, and not to a unique, stochastic outcome (for this reason these calculations are supplemented with interpretations that account for this stochasticity, see Sections \ref{sec:first} and \ref{sec:different}); third, decoherence calculations cannot do without including randomness from the onset, they only move it back to the initial state which is assumed to be described by a set of sufficiently random real numbers, leading to sufficiently uncorrelated states at later times. But "real numbers are not really real" \cite{gisin2018indeterminism}, they are only mathematical objects. Infinite precision is not a physical concept \cite{infinity}. And finally, due to the emission of thermal radiation a heat bath is not a closed system and cannot be turned into a closed system by including a finite, larger environment. In contrast, our description of the heat bath has none of these problems as it is based on empirically known facts, and openness and stochasticity are explicitly included as its features. In the next subsections, we will see how it gives unique outcomes for the electron time evolution.

\subsubsection{Unraveling the time evolution of the reduced density matrix}
\label{HB_unravel}
With the outlined understanding of the nature of a heat bath, the density matrix (\ref{rhoend2}) cannot be interpreted as the reduced density matrix of a pure state that represents an entanglement of the system and the heat bath. Only the electron can be described by a wave function, not the combined system. Due to the limited range of quantum coherence of the heat bath, the electron wave function must become localized within the distance of this coherence length. This means that not only the phonons, but also the electron is described by wave packets of a limited extension. 

In order to obtain an equation of evolution directly for the electron wave function, the theory of open quantum systems \cite{breuer2002theory} provides useful tools. These tools allow the unraveling of the time evolution of a density matrix, which follows a Lindblad equation, in terms of a stochastic time evolution of a wave function. Such an unraveling can be done in different ways, by introducing jump processes or continuous-time stochastic evolution \cite{gisin1992quantum}.  We suggest that a continuous-time stochastic wave function dynamics is indeed the correct way that the density matrix that describes the electron should be interpreted. In fact, all calculations that rely on semiclassical dynamics of wave packets presume that a continuous localization process occurs, which is nonunitary and has stochastic contributions. 

The time evolution of the density matrix leads to   an important feature of the wave function trajectory: Since the density matrix evolves to a blockdiagonal form, it does not allow transitions of the wave function between the different Hilbert spaces that are represented by the blocks.  Thus the time evolution of the electron wave function is non-ergodic and gets trapped in one of the diagonal blocks! This is exactly what is needed in order to obtain a "collapse" of the initially entangled state of the electrons that interact with the photon to a non-entangled state in which one of these electrons (or none) ends up in the conduction band. It is also needed for the above semiclassical and classical description of the subsequent chain of events, until the detector indicates the measuring result.  

\subsubsection{Feedback  between quantum system and heat bath}\label{HB_feedback}

We have mentioned above that an important feature of a heat bath is the random exchange of energy between its degrees of freedom. This means that there are feedback loops between these different degrees of freedom. Such feedback loops introduce nonlinearities into the system, and this cannot be described by the unitary time evolution of a wave function \cite{ellis2012}. 
Unitary time evolution excludes feedbacks, as is best visible in QFT calculations, where a wave function propagates from its source (or initial state that is prepared by a nonunitary process similar to the measurement process) to its final state, where it is measured or absorbed (again a nonunitary process).  

We have taken into account these feedbacks or nonlinearities implicitly by arguing that the time evolution of the heat bath is not unitary. However, we have not taken into account feedbacks between the electron and the heat bath. 

In this respect, we did not yet go sufficiently far away from decoherence theory. We have used a Lindblad equation, which is linear in the density matrix and  does not include an influence from the system on the environment, although the presence of the electron clearly affects the dynamics of the surrounding nuclei. Admittedly, there does not yet exist a theory that includes this effect, and deriving it would define a research program in its own right. Nevertheless, theoretical tools of the type required here are available to some extent: once the effect of the electron on the heat bath is included,  there is a feedback loop from the electron via the heat bath back to the electron, which means that one obtains a nonlinear Schroedinger equation. Here is a simple model that implements this idea: we write the Schr\"odinger equation of the electron 
as
\begin{equation}
i\hbar \frac{\partial \psi}{\partial t} = (T + V)\psi
\end{equation}
with the potential $V$ being due to the external environment, which is formed by the ions (screened by the presence of other electrons) of the lattice of the detector material. The potential itself changes due to the changing electron wave function, which exerts a force on the ions that depends on the electron charge density,
\begin{equation}
\frac{\partial V(\vec x,t)}{\partial t} = f(\vec x,[|\psi|^2])\, .
\end{equation}
The functional $f$ depends in general on the entire function $\psi(x,t)$. Integrating this equation and inserting it into the equation for the electron wave function gives
\begin{equation}
i\hbar \frac{\partial \psi}{\partial t} = \left(T + \int_{-\infty}^t dt'f(\vec x,[|\psi(t')|^2])\right) \psi\, .
\end{equation}
This is a nonlinear Schr\"odinger equation that is nonlocal in time and space. If we assume that the electron wave function changes slowly, the time integral on the right-hand side can be replaced with $f([|\psi(t)|^2])$. A particular example of such an equation is the Schr\"odinger-Newton equation, which is a Schr\"odinger equation where the potential term describes gravitational self-interaction of the wave function. This equation has been suggested to describe gravitational wave-function collapse \cite{diosi1984gravitation,penrose1996gravity}, but cannot be the full explanation since it is deterministic \cite{bahrami2014schrodinger}. When the equation is simplified further by including only a local interaction, one arrives at the well-known class of nonlinear Schr\"odinger equations, which display a wealth of different phenomena, for instance solitons. 

All this shows that by including a feedback with the environment, a localization of the wave function can in general be obtained. A full and satisfactory theory, which includes also stochastic effects due to the fluctuations inherent in the thermal motion of the ions, would still need to be worked out.

\subsubsection{The heat sink}\label{HB_heatsink}

Decoherence theory assumes that the heat bath (or environment) can be a closed system with a unitary time evolution. 
However, the true environment within which the measurement process takes place is much larger than this. We have already mentioned that a heat bath emits thermal radiation into the surrounding world. The heat sink is additionally required for the reset process. 

In fact, the wider environment provides both sources for incoming particles that deposit high-grade energy in the system (such as the photon that is being detected by our setup), and a heat sink that allows irreversible processes to take place by freely absorbing low grade output radiated energy, i.e.,  the heat sink is essentially unaffected by the energy sent to it. The ultimate heat sink for the irreversible processes on earth is the dark night sky (Penrose \cite{penrose2162emperor}:413-415; \cite{penrose2016fashion}:256-258). It receives  radiation that has a higher entropy (per amount of energy) than the radiation that originally reached the earth from the sun.  This energy flow to and from the earth is what creates an overall non-equilibrium situation  and so allows irreversible processes, such as the measurement process, to take place. 
Ultimately the heat sink exists because of the very special initial state of the universe which \textit{inter alia} leads to a dark night sky that can receive our low grade waste radiation (Penrose \cite{penrose2162emperor}:391-449). The thermodynamic arrow of time depends on these special initial conditions. This is discussed further in Section \ref{sec:arrow} below.

\subsection{Causality}\label{sec:problems-causality}

It is often argued that the concept of causality has become superseded, since it is now replaced by functional relationships that relate initial and final states by deterministic equations of motion, which are furthermore time-reversible \cite{norton2009there}. The reasons for this view are that the supposedly fundamental theories of physics are deterministic and time-reversible. However, this view is based on a rather incomplete and idealized view of physics \cite{causality}. 
When considering the various steps involved in the 
calculations outlined in previous sections, it becomes clear that this simple view is not sustained in the calculations that are performed in practice. In particular, one assumes that the initial state of the measurement device is in a (metastable) equilibrium, which is affected neither by previous measurements, nor does it know yet about the incoming photon. 
This initial state is often the result of a reset process after the previous measurement, see the discussion in Section \ref{sec:reset}. It is important that this state is reproducible, so that the changes that will be observed during a measurement can be said to be caused by the incoming photon. 
Note that using the words ``initial state'' has already assumed the arrow of time, which underlies the usual notion of causation:  an effect occurs after its  cause.

The concept  of an initial state  that does not carry unwanted or uncontrolled traces of past events is an essential prerequisite of the linear response theory calculation that is used for describing electron conduction: the system starts out in thermal equilibrium, and the step function $\theta(t-t')$ in (\ref{chiAB}) explicitly breaks time reversal invariance.  The Boltzmann equation (\ref{BE}) and the avalanche equations (\ref{avalanche}) also break time reversal symmetry and implement the idea that a ``cause'' (here: a moving electron) triggers an "effect" (here: the moving of more electrons into the conduction band). Any more detailed, quantum-mechanical description of this process would need to employ Lindblad-type equations such as (\ref{eq:Lindblad}), which are again not invariant under time reversal.   As we have discussed thoroughly in Section \ref{sec:problems-heat_bath}, all this is enabled by the truly stochastic nature of a heat bath, which allows to forget the past and to establish an equilibrium state that is determined by nothing more than the macroscopic variables that control it. 

\subsection{Classical ingredients}\label{sec:problems:classical_ingredients}

The above calculations and considerations demonstrate the fact that quantum calculations and quantum processes always depend on a classical context \cite{drossel16quantumclassical}.  Discussions of quantum experiments certainly do so: they rely on many classical components, including  classical detectors, as can be seen in any experimental quantum paper. For example \textit{Diagram 1b} of \cite{gisin2017experimental} shows many classical devices including  single-photon detectors,  a periodically poled KTP waveguide,  monochromatic laser,  dichroic mirror, a fiber-based 50/50 beamsplitter,  electro-optical modulator,   polarization beamsplitter, and the quantum memory  itself.  

In more detail as regards the detector, its classical structure determines the possible processes that can occur. In our example, the structure is designed such that there are bound electron states that can make a transition to conduction electrons upon excitation due to the interaction with an incoming photon. The macroscopic structure defines the degrees of freedom that constitute the internal heat bath. It also defines the nonequilibrium conditions that are exploited when removing the electron from its original site via an electrical field and when the avalanche process takes place. The detection process itself involves a definite transition to a (classical) final state of the apparatus that is different from the initial state. 
And above all this, the entire description of a quantum process presupposes the concepts of space and time, which are also classical concepts.

While many researchers aim at deriving the classical world from the quantum world (e.g. \cite{hartle2011quasiclassical}), our view is different: we agree that  the classical world emerges from the quantum world in the sense that the world is quantum on small scales everywhere. However, this does not determine the classical world bottom-up since the quantum description has its limits of validity. When you knit the local quantum bits together, you 
get a world that can no longer as a whole be described in quantum terms \cite{ellis2012}. This is related to a limited applicability of unitary time evolution when dealing for instance with heat baths \cite{drossel2017ten}, as is carefully described above (Section 5.2). 

Very generally, a fully quantum mechanical, unitary time evolution of a quantum system is always confined to limited time interval between ``preparation'' and ``detection''. It takes place within a given Hilbert space that is always defined by the macroscopic context in which the time evolution occurs. Often,  potentials $V(\x)$ are included in the Schr\"odinger equation, which means that the external environment is treated classically and not as a set of quantum particles that are entangled with the considered system. 

In contrast, a detection process, and in fact any quantum event, is accompanied by a change in the classical world, i.e., in the world that defines the degrees of freedom and the environment for quantum processes. It involves a change of Hilbert spaces, not dynamics within given Hilbert spaces.

The classical world already has an arrow of time built in, associated with special macroscopic initial conditions for the universe (a ``Past Condition'': \cite{sciama2012unity,ellis1972global,albert2001time},  \cite{penrose2162emperor}:391-449, \cite{EllisTopDown}:280). Thus it is plausible that the microscopic arrow of time occurring in the quantum collapse process \cite{ellis2012}  is derived by top-down action from the macroscopic direction of time set by cosmology. This is discussed further in Section \ref{sec:arrow}.

\subsection{Top-down effects from the wider context}\label{sec:top_down_effects}

Overall, this paper is an example of the way that top-down effects work in physics \cite{EllisTopDown}. There are top-down effects from the macroscopic setup of the measurement; from the more detailed, microscopic internal structure of the measurement apparatus;  from the wider world, which in particular is responsible for the arrow of time; and from the mind of the experimenter, which determines what measurement will take place and when.

{\it At the macro level}, the measurement setup determines what kind of measurement is done. In our example, the detector is designed for the detection of photons and can be used in the context of position measurements, as illustrated by the two-slit setup. Very generally, measurement devices are constructed and set up in a way that the  observable of interest, such as energy, polarisation, charge, etc.~can be measured. They can be employed in various ways, for instance by choosing the directions of polarisation to be measured in a polarisation measurement, or by deciding whether a which-way measurement is attempted or not in a 2-slit experiment, and so on.  

{\it At the micro level}, the context for the interaction is set by a specific metastable structure that allows transitions upon impact of the particle that is to be detected. In our example, this microscopic structure is represented by bound states and by a continuum of states (conduction band) of the electrons. This is provided by the semiconductor material and  leads to a work function for freeing up an electron,  depending on the nature of the material (hence a top-down effect \cite{ellis2012}). The  structure of the crystalline context (semiconductor material) furthermore leads to a macroscopic number of internal degrees of freedom, which constitute a heat bath of quasi-particles (phonons), which plays a crucial top-down role at localizing the electron. Similarly, the other examples for measurement devices given in Section \ref{sec:building_blocks_context} have their specific microscopic internal structure that is set up in a metastable state and that provides also a heat bath. 

{\it At the global level}, 
the external heat sink provides the arrow of time that is integral to the process, ultimately originating in the expansion of the universe and the associated Direction of Time, as we discuss in Section \ref{sec:arrow}. 

{\it At the mental level},  top-down effects due to human decisions and actions determine what apparatus is used to measure what and how it is designed, and so causally underlie the existence and nature of the experimental apparatus, and hence play a key role in its outcomes. Furthermore an additional top-down effect comes from the context that produces the particle to be measured, which in some instances is also decided by the experimenter (for example the particles detected by the LHC at CERN were produced by a very complex set of accelerators designed for that purpose by engineers). These are both aspects of the causal power of thoughts \cite{EllisTopDown}, enabled  by the mind-brain relation \cite{gabriel2017not}.

\subsection{The arrow of time}
\label{sec:arrow} 
The  fundamental issue of the arrow of time has been a recurring theme in previous sections: how does the irreversible nature of the measurement process, with an associated arrow of time, arise out of the time-symmetric underlying unitary dynamics embodied in the Schr\"{o}dinger  equation, which has no arrow of time? In broad terms the answer has already been given: it is determined in a top-down way by the environment (Section \ref{sec:top_down_effects} and  \cite{EllisTopDown}). But more precisely, how does that work? 

The proposal here will be that this works in three stages. First, we live in an evolving block universe that sets the global \textit{direction of time} (Section \ref{sec:EBU}). Second, this is tied to a limited information content and stochastic transitions, which generate together a \textit{thermodynamic arrow of time} that is aligned everywhere with the cosmological arrow of time (Section \ref{sec:Therm_arrow}). Third, the thermodynamic arrow of time prevents the conversion of heat into a coherent emission of electromagnetic radiation, leading to the  \textit{electrodynamic arrow of time} that necessarily  agrees with the direction of time set by cosmology (Section \ref{sec:Elec_arrow}).

All this then sets the context whereby heat baths can communicate the macroscopic arrow of time to quantum systems, as discussed above, thus determining the quantum arrow of time (as experienced in the measurement process). Thus the arrow of time in quantum measurements in the end derives from cosmology \cite{sciama2012unity,ellis1972global}: a crucial top-down effect (Section \ref{sec:top_down_effects}).

\subsubsection{The Evolving Block Universe}\label{sec:EBU}
 The idea here is that we live in an Evolving Block Universe (EBU) \cite{ellis2006EBU} \cite{ellis2014evolving} \cite{ellis2014EBU} which sets the direction of time as a global quantity, determined by cosmology as understood today \cite{dodelson2003} \cite{EMM}\cite{peter2013primordial},  to be contrasted with various arrows of time (e.g. electrodynamic and thermal), which are quantities related to local physics, as discussed below.
 
 The point is that the Einstein Field Equations determine the evolution of 4-dimensional space time ${\cal M}$ from initial data, but do not specify the boundaries of that space time. In the kind of evolving universe in which we live, the start of the universe is represented by an initial singular boundary ${\cal B}_-$. We can represent the present time as a spacelike surface ${\cal P}(t)$  which is situated a proper time $t$ after the initial singularity, as measured along the fundamental world lines of cosmology. At each instant $t$ is the age of the universe, which in 2015 was measured by the Planck team to be $t_0 = (13.813 + T_0) \times 10^9$ years, where $|T_0| <0.038$. The present time keeps increasing, as time passes; at the time of writing in 2018, the age of the universe is $t_0^* = t_0 +3$. Thus the future boundary of spacetime keeps moving to the future as time progress. So at a time $t$, the spacetime ${\cal M}(t)$ contains all events from ${\cal B}_-$ (at $t=0)$ to ${\cal P}(t)$, but none to the future of  ${\cal P}(t)$. At 
 time $t+\Delta t$ ($\Delta t > 0)$ the spacetime ${\cal M}(t+\Delta t)$ contains the whole previous spacetime ${\cal M}(t)$ plus the small extra bit $\Delta {\cal M}(t)$ lying between the surfaces ${\cal P}(t)$ and ${\cal P}(t+\Delta t)$.   
 
 The viewpoint then is as follows: at time $t_0$ the past 4-dimensional cosmological region ${\cal M}(t_0)$ exists, because it contains events that causally affect what happens at ${\cal P}(t_0)$ today (for example, nucleosynthesis occurring inside stars in the past provided the elements out of which planets are made today). The present ${\cal P}(t_0)$ exists: it is the set of events where the indefinite future gives way to the definite past. The future of  ${\cal M}(t_0)$ does not yet exist because it has not yet been determined what will happen there, and events that will happen there (when it comes into existence) cannot reach back to the present day to change what happens today (at a macro scale). Thus the future is a space of partially determined possibilities (conservation laws have to be obeyed) that will come into existence in the future. It does not actually exist at the present time. The spacetime as a whole is a 4-dimensional block ${\cal M}(t)$ whose future boundary keeps extending into the future as time $t$ progresses.
 Space time exists  at time $t_0$ from $t=0$ to $t =t_0$ but not at later times \cite{ellis2006EBU,ellis2014EBU,ellis2014evolving} (see Figure 1) (note that while we have discussed this in the context of the standard Friedmann-Lema\^{i}tre models of cosmology, the same structure will hold much more widely, for example in perturbed cosmological models \cite{dodelson2003}). 
 
  Finally the \textit{direction of time} is the globally determined arrow pointing from the initial time $t=0$ at the fixed starting boundary  ${\cal B}_-$ of the EBU to the ever moving future boundary ${\cal P}(t)$, which is the present.

\subsubsection{The thermodynamic arrow of time}
\label{sec:Therm_arrow}
So far, we described the concept of an EBU within the context of Einstein's field equations. This is a coarse-grained description. Implicitly hidden in the EBU concept is the idea that the present state, if described microscopically, does not contain sufficient information to fully fix the future time evolution, which \textit{inter alia} is implied by the standard quantum uncertainty relations applied to initial data at any time $t_{init}$. This means that unitary time evolution has a limited applicability, and that stochastic events must occur that decide which of several possible states will be realized, 
or provide an ensemble of possibilities from which top-down selection processes can decide an outcome on the basis of higher level selection criteria \cite{EllisTopDown}.

This agrees with a realist interpretation of statistical physics: The entropy of a system is, apart from a factor $k_B \ln 2$, identical to the number of bits required to specify the microstate of the system, given its macrostate. If we take this seriously, the state of a thermodynamic system is not defined more precisely than with this number of bits. Accordingly, transitions between microstates are stochastic within the formalism of statistical physics. They are not determined uniquely by the underlying quantum physics for reasons discussed in  section \ref{sec:problems-heat_bath} on heat baths. From a discrete set of states, combined with stochastic transitions, an increase of entropy
in the future direction of time
follows in a natural manner, since time evolution leads locally to those macrostates that can be realized in  many  
more ways by microstates because they occupy vastly  larger regions of phase space \cite{penrose2162emperor}. In an isolated system, the time evolution leads to an equilibrium which has maximum entropy, provided the system is ergodic, i.e.~that for each initial state there is a time after which the system can be in any of its states with a nonvanishing probability. In our discussion of the heat bath above, we have made explicit use of the stochastic nature of systems in thermodynamic equilibrium. 

We have thus argued that the thermodynamic arrow of time is inseparably linked to the cosmological arrow of time, where the expansion of the universe  leads to a decrease of the temperature of the primordial plasma as the time passes and the standard thermal history of the universe, involving baryosynthesis, nucleosynthesis, decoupling,  and so on,   ensues \cite{dodelson2003,EMM,peter2013primordial}. Both arrows of time are tied to a limited amount of information specifying the present state of the universe and stochastic transitions that realize one of several possible future states. 

In order to obtain the nonequilibrium situation realized in the universe, it is necessary that local entropy is not already maximised, which is a \textit{Past Condition}, as discussed by Albert \cite{albert2001time}. That requires, as pointed out by Penrose (\cite{penrose2162emperor}:391-449; \cite{penrose2016fashion}:241-258), that the universe must start off in a very special initial condition where inhomogeneities are limited (else the huge possible entropy of black holes would imply a highly inhomogeneous state of existence with many primordial black holes and no development of galaxies and planets).  The expansion of the universe leads to an increasing number of possible states, and it is responsible for maintaining a  nonequilibrium situation, which relaxes locally only very slowly, due to various "hangups" created by gravity, as explained beautifully by Dyson \cite{dyson1971energy}.

\subsubsection{The Electrodynamic and QFT arrows of time}\label{sec:Elec_arrow} 

The emission of electromagnetic radiation plays an important role in the irreversible cooling of finite-temperature systems, and we have included this process in our discussion above. We therefore turn now to the electrodynamic arrow of time. 

One can in principle solve the electrodynamic equations via a Green's function method using either a retarded Green's function integrated over the past, an advanced Green's function integrated over the future, or a combination of the two \cite{wheeler1945Feynman}. The results of these different types of calculations are the same if the initial or final conditions and boundary conditions are chosen in each case in such a way that they describe the same physical process. 
 
Nevertheless, the physical processes that produce and absorb electromagnetic radiation are not invariant under time reversal. 
The emission of radiation from an oscillating charge distribution occurs into the future, not into the past, and this is why the advanced potentials are discarded and the retarded potentials are retained in calculations. The mathematical description of the emission of radiation by using advanced Green's functions requires boundary and final conditions that are highly correlated in a nonlocal way because they 
represent the future state of the electromagnetic fields that are emitted from the source and that radiate away to infinity (or until they are absorbed). They are therefore of no help. Electrodynamics thus has an arrow of time that indicates that electromagnetic fields originate from localized sources in the past. This is the Sommerfeld radiation condition. A good discussion of this issue can be found in \cite{weinstein2011electromagnetism}.
The local emission of radiation into the future is related to the EBU: local events produce photons that can causally affect future events, but not past events. This helps us to specify the properties of "events" in the context of the EBU: Events must occur locally within a limited region of space. Their causes are other (local) events that occurred in the past, and they can influence later events that will take place in their forward light cone. In this way, the electrodynamic arrow of time follows directly from the cosmological one.

The electrodynamic arrow of time is furthermore closely connected to the thermodynamic arrow of time: walls and clouds and other extended objects absorb radiation and transform the absorbed energy into stochastic thermal motion. They don't use heat in order to emit radiation in a coherent way such that this radiation converges on "sources" that lie in the future: this would be the time-reversed process to emission of radiation by an oscillating charge, and such `pre-established harmony' is disallowed by the Law of Conditional Independence (Penrose and Percival \cite{penrose1962direction}). They emit thermal radiation that has a limited coherence length. The emission of photons from different surface sections is uncorrelated. Both of these processes, the absorption of coherent electromagnetic waves  and the emission of thermal photons, increase entropy and thus obey the second law of thermodynamics.  

  However there is a further important point. Thermodynamics establishes that a heat bath can only radiate energy in this way if the environment is at a lower temperature than the heat bath: else it is heated up by absorbing thermal radiation from that larger environment, rather than being able to cool by radiating energy to it.  We therefore need special initial conditions in the Solar System, establishing a dark night sky, in order that the heat bath can in fact cool by radiating heat in the future direction of time. The temperature of interstellar space at the present time is $2.725K$ - the present temperature of the Cosmic Background Radiation that was emitted at the surface of last scattering in the early universe at a temperature of $4000K$ and redshift of $1100$ \cite{dodelson2003,EMM, peter2013primordial}. This low temperature  of the night sky, allowing the Earth's ecosystems to function in thermodynamic terms, is a result of the initial state of the expanding universe \cite{penrose2162emperor}.

\begin{figure}
\centering
\includegraphics[width=0.5\textwidth]{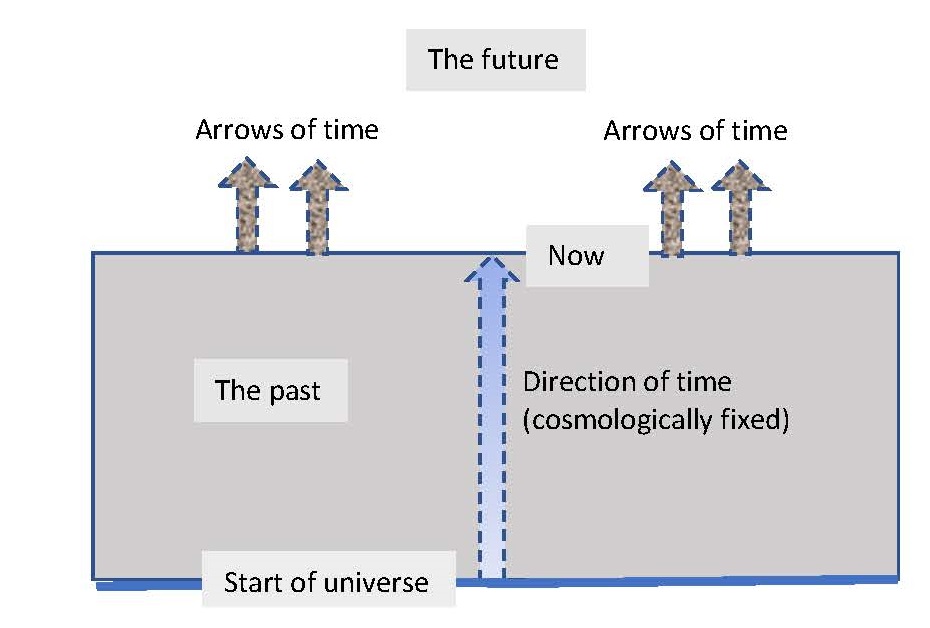}
\caption{\it{The distinction between the global} Direction of Time, {\it set by the expansion of the universe since its start, and  local} Arrows of Time, {\it determined by local physical properties \cite{ellis2014evolving}.  In the expanding universe, the start is a fixed initial boundary to spacetime. The present is an ever changing future boundary to spacetime: as time passes it moves to the future, in the  direction indicated by the Direction of Time. At each instant spacetime exists from the start up to the present, but not to the future (because of quantum randomness, it is not yet determined what will happen in the future,  so it is not a spacetime domain with definite properties). Local Arrows of Time include quantum, electrodynamic, gravitational,  thermodynamic,  biological, and psychological}}.    
\end{figure}

Similar considerations are applied when solving QED and QFT equations formulated via Feynman propagators \cite{feynman1965quantum,peskin2018introduction} - they too are integrated using retarded Green Functions. This reflects the idea that the unitary time evolution of a quantum particle originates in a local source that lies in the temporal past. Thus in this way there is a direct effect of the direction of time in terms of establishing an arrow of time for quantum field theory.  

With the local thermodynamic Arrow of Time established in the same direction as the electrodynamic Arrow of Time, both being aligned with the cosmological Direction of Time, 
this then chains down \cite{ellis2014evolving} to provide the arrow of time in the various steps in the quantum measurement process (Section \ref{sec:model-steps}) via the properties of the heat bath, as discussed above (Section \ref{sec:problems-heat_bath}). 
Thus the quantum Arrow of Time is a particular case of the  contextual effects discussed in Section \ref{sec:top_down_effects}. The biological and psychological Arrows of Time follow from the others.

\section{Discussion}\label{sec:discussion}
In this section we summarize the main achievements and new ideas in this paper; point out how one might be able to explore the intermediate regime that is neither quantum nor classical; and explain the theoretical challenges or possibilities to deepen and expand the many issues that arise.

We first discuss what we have done (Section \ref{sec:done});  then the difference between our  approach and other similar ones (Section \ref{sec:different});
the difference between events and measurements (Section \ref{sec:discussion_comments}), leading to consequent possible further applications of our approach; and finally some issues arising from what we have done (Section \ref{sec:outcomes}).

\subsection{What we have done}\label{sec:done}
We have outlined a comprehensive approach to the quantum measurement process, with the following properties:
\begin{itemize}
\item We derive an effective collapse dynamics (\ref{collapse}) from the unitary dynamics (\ref{unitary}) of standard quantum theory. 
We do not introduce any \textit{ad hoc} terms into the equations to effect this result. Rather they come about because we take into account the context within which realistic measurement processes take place \cite{ellis2012}. It is crucial for our treatment that this context is not quantum but classical. 
\item The macroscopic, classical measurement apparatus  determines by its structure the possible types of measurement events. This structure permits a cascade of events from the microscopic to the macroscopic scale such that a measurement result can be read off by an experimenter. Furthermore, this structure provides a macroscopic number of internal degrees of freedom that create the local heat bath that interacts with particles during detection. 
\item The bridge between the quantum and classical world is formed by the heat bath, which is characterized by a limited temporal and spatial range of unitary time evolution and by stochasticity \cite{drossel2017ten}. A description by wave functions is therefore only locally possible, as wave packets with a limited scope for entanglement. This leads to a "collapse" or localization of the wave function of the electron that has been lifted to the conduction band by the interaction with the incoming photon. 
\item We provide physical equations that describe each step from the initial interaction of the incoming photon with the semiconductor material to the final deflection of the pointer, by borrowing from the most suitable theories available in the literature. Apart from the very first step, none of these steps is described by deterministic, unitary time evolution of a wave function. This underlines our view that unitary quantum evolution has a limited scope.
\item By emitting thermal radiation the heat bath is coupled to the wider environment. We do not model explicitly this wider environment,  but it is important for our arguments as to why the heat bath cannot be described by an overall wave function.  This wider environment provides a heat sink and thus enables the irreversibility that appears in the overall process, overall tracing back to the evolution of the universe. A further irreversible step is the reset process of the detector that prepares it for the next detection event.  
\end{itemize}

\subsection{Comparison with other approaches}\label{sec:different} Our approach has benefited a lot from the achievements of the Copenhagen interpretation, of decoherence theory, and of stochastic collapse theories. Furthermore, it shares with consistent histories the view that stochasticity is fundamental and that the calculations to be performed are similar to those of decoherence theory. In the following, we explain in which respects our view agrees with and differs from these types of approaches.
\begin{itemize}
\item \textit{Copenhagen interpretation}: The Copenhagen interpretation takes explicitly into account the classical world and its  top-down effects on the quantum world. A measurement always requires a classical measurement apparatus or observer, and during a measurement a "collapse" of the wave function occurs, which is a stochastic, nonunitary process. We fully agree with this description and hold that it cannot be reduced to, or replaced with some other, supposedly more "fundamental" theory. Neither do we believe that it requires the acceptance of a non-realist metaphysical  position, such as an instrumental, positivist, or pragmatist view. Rather, we propose that quantum and classical physics both have their range of applicability, and that neither can be reduced to the other. The bridge between these two regimes is formed by the heat bath, which obeys unitary time evolution only over limited time and length scales that depend on temperature, as we have discussed extensively.  Of course, this raises the questions of how to explore and specify the properties of this transition range between the quantum and classical world, and how to relate this to the experimental observation of entanglement of huge numbers of particles or over large distances.   We will address these questions further below.
\item \textit{Decoherence theory}: An important achievement of decoherence theory lies in the demonstration that the coupling to an environment with many degrees of freedom causes an evolution of the reduced density matrix of the quantum system from a pure to a mixed state that, under suitable conditions, gives the correct probabilities for the different possible measurement outcomes.  We have made use of this type of calculation when describing the interaction of the electron with the heat bath, which makes the transition of the electron to the conduction band definite and localizes it in the conduction band. In contrast to decoherence theory, we do not think that the environment that is relevant for the measurement process can be described by unitary time evolution. In the theory of open quantum systems, one starts from a unitary time evolution of the combined quantum system and environment and obtains, by making suitable approximations and assumptions, the time evolution of the reduced density matrix. This poses  the threefold problem that decoherence cannot explain individual measurement outcomes, irreversibility, and stochasticity. In Section \ref{HB_decohere} we have explained that these three problems are resolved by abandoning the hypothesis that the heat bath follows unitary time evolution. 
\item \textit{Consistent histories}:
The consistent  histories interpretation \cite{griffiths1984consistent,gell1993classical} is based not on wave functions, but on stochastic histories which are calculated using projection operators and decoherence functionals. It is the furthest developed of those interpretations of quantum mechanics that are based on the standard formalism, and it has the appealing property of including stochasticity from the onset. The calculations are mathematically very similar to the calculations of decoherence theory. As mentioned in  Section \ref{HB_decohere}, we consider decoherence calculations as not fully satisfactory, as they must be supplemented by assumptions of uncorrelatedness or randomness of environmental degrees of freedom. Another difference of our view from consistent histories is that consistent histories are applied to closed systems and  are reversible in time, while we consider top-down effects from the classical world, the coupling to a heat sink, and the temporal irreversibility due to the emission of thermal radiation.
 These three features are in our view essential for establishing definite outcomes of stochastic events and an objective reality. We regard these as key desirable achievements of our approach that cannot be achieved by competing theories. In particular, this  cannot be achieved with the consistent histories approach, which leads to several possible frameworks (with the associated framework-dependent 'true' and 'false' statements). 
\item \textit{Stochastic collapse theories:} These theories postulate stochastic corrections to the time evolution of the Schr\"odinger equation such that the wave function becomes localized in space or with respect to some other relevant observable \cite{bassi2013}. These stochastic corrections do not depend on context but occur universally for each particle. Instead of using a modified Schr\"odinger equation, the description of the time evolution can also be given in the ensemble picture, using density matrices. The density matrix equations that we mentioned in the previous paragraph, which are obtained for the time evolution of open quantum systems, represent in fact also stochastic collapse models. They can be unraveled in terms of the stochastic time evolution of a wave function. We have argued that this is the correct way to interpret the time evolution of the density matrix that we obtained. In this sense our description involves also a stochastic collapse model. However, the type of collapse equation that we use depends on the local context, as we start out with the linear superposition of electron states generated by the incoming photon and couple this to the heat bath. So in our view, the type of collapse equation to be used, and the conditions under which it is to be used, is determined top-down by the local classical context.
\end{itemize}
An interesting other class of collapse equations is given by gravitational collapse theories \cite{penrose1996gravity}. These  depend also on context, but through gravity, hence it is then not the {\it local} context that matters, as in our case (the gravitational effect of the measuring apparatus on the interacting photon and electron is negligible). Another interesting variant is the Montevideo interpretation of quantum mechanics, which consists in supplementing environmental decoherence with fundamental limitations in measurement stemming from gravity \cite{gambini2015montevideo}. Interestingly, there are various connections between gravitational theories and thermodynamics, and we do not want to rule out that our views concerning the heat bath may be relevant also for the huge number of degrees of freedom occurring in theories that relate quantum physics to gravity. 

Furthermore, there exist interesting suggestions to introduce causality and irreversibility into physics at a fundamental level via causal sets \cite{sorkin1990spacetime, cortes2014quantum}. Again, the route to relating this approach to real-life measurement processes with their top-down effects is not clear. The view that is probably closest to ours is the contextual approach by Grangier and Auff\`eves \cite{grangier2018quantum}.

\subsection{Events and Measurements}\label{sec:discussion_comments}

A key distinction we made at the start (Section \ref{sec:building_blocks_context}) was the distinction between `events' and `measurements'.  The latter are rather rare special cases of the former, which occur whenever a wave function projection (\ref{collapse}) takes place. Measurements involve the building blocks mentioned in Section \ref{sec:building_blocks_context}, since the detection system must be constructed such that a measurement result can be read off by an experimenter. 
State vector preparation is very similar to measurement: It involves macroscopic devices in order to generate an eigenstate where there was not one before. 
It occurs either by selective absorption or by separation and collimation \cite{ellis2012}.

We consider here the ubiquity of events (Section \ref{sec:events_ubiqu}), the nature of events (Section \ref{sec:events_nature}), and further examples of events (Section \ref{sec:events_further_examples}). 
\subsubsection{Ubiquity of events}\label{sec:events_ubiqu} Beyond these specific state preparation and state measurement scenarios, events take place all the time everywhere. While they are not measurements according to our definition, they are  nevertheless characterized as `measurements' in standard texts such as Dirac \cite{Dirac1982}. Throughout this paper, we have referred to events at various places. In fact, we have presumed that a huge number of events happen during a single measurement: 
\begin{itemize}
\item Initially, the interaction of the detector material with the incoming photon produces a linear superposition of an excited electron and electrons in the ground state, which then collapses to a definite outcome with the photon being either absorbed or not absorbed and at most one electron being in the conduction band, see Section \ref{sec:problems-collapse}. 
\item While the electron moves in the conduction band, it is described as a localized wave packet. However, in order to remain a wave packet, it must undergo repeated collapse events that localize it. 
\item During the avalanche process, additional electrons are kicked into the conduction band. Each of these transitions to the conduction band is also an event, and all the conducted electrons remain localized by further events. 
\item The rotation of the coil and associated pointer deflection is a classical, macroscopic change. Such classical changes must be associated with a series of events that localize coil and pointer (and all other classical parts of the detection device) and prevent superpositions. 
\item The reset process of the detector involves a whole series of events, with a complexity similar to that of the series that we just described. 
\end{itemize} 
Many further examples are given in Section \ref{sec:events_further_examples}.
\subsubsection{The nature of events}\label{sec:events_nature} At various places of this paper, we have mentioned  requirements for events and the character of events: 
\begin{itemize}
\item The interaction with a heat bath with a macroscopic number of degrees of freedom is crucial for the occurrence of an event. We defined a heat bath in a much narrower sense than is often done in the quantum community, see Section \ref{sec:problems-heat_bath}.
\item An essential feature of the heat bath is that its microscopic state cannot be controlled from outside and that it cannot be repeatedly prepared in the same microscopic state. Its state is specified with no larger precision than that implied in the entropy formula.
\item The heat bath emits thermal radiation, i.e., it undergoes irreversible changes. This irreversible emission of radiation makes the events that occur in contact with the heat bath also irreversible. Furthermore, the emission of these thermal photons is of course also a series of events. 
\item The last two points imply that a heat bath is not an object that can be described by a wave function that undergoes unitary time evolution. If collapse events are to happen, it is  necessary that there are systems that do not evolve unitarily.  
\item Just as we have discussed above (Section \ref{sec:top_down_effects}) how measurements always take place in a way shaped by the local context, the same is also true for events. The combination of higher level contexts sets the stage for the possible specific outcomes (\ref{collapse}), as has been demonstrated explicitly in the previous sections for a specific case.
\item Measurements  and other events act as a dynamics {\it of} Hilbert spaces, in contrast to unitary time evolution given by the Schr\"{o}dinger equation, which acts as dynamics {\it within} given Hilbert spaces. Quantum theory {\it per se} does not tell us what Hilbert spaces to use. This requires the classical, macroscopic context. Our paper therefore emphasizes repeatedly the importance of top-down effects from the macroscopic context. Changes of this context cannot be achieved by unitary time evolution but require events.  
\item All this is closely associated with the idea of an evolving block universe \cite{ellis2014evolving}. This idea requires that the initial state of the universe does not yet specify the details of its future time evolution. Instead, definite, irreversible events really happen and time really passes, see section \ref{sec:arrow}. 
\item The entropy of the universe increases with each event as photons are emitted from the heat bath to a colder environment and ultimately to the dark night sky, which represents a heat sink. 
\item Occurrence of events, including quantum measurements, are cases of symmetry breaking whereby emergent phenomena have different properties than the strata out of which they emerge, as explained clearly in the classic paper ``More is  Different'' by Anderson  \cite{anderson1972more}. 
Thus the top-down effect of the local context  in a measurement (Section \ref{sec:top_down_effects}) breaks both the time symmetry of unitary evolution (Section \ref{sec:arrow})  and Lorentz invariance (the measurement apparatus necessarily has a preferred rest frame in which the experiment takes place). The same will be true for all events: the local context will also break Lorentz symmetry.
\item This view of events is closely tied to the concept of causality,  which requires a partial ordering between events \cite{causality} and the possibility of state preparation, see also Section \ref{sec:problems-causality}.
\end{itemize}

\subsubsection{Further examples of events}\label{sec:events_further_examples} There are many other types of events that happen in some specific context, for which our contextual approach is relevant. Here are some examples from condensed matter physics, biological physics, and cosmology:
\begin{itemize}
\item The growth of a crystal in a supersaturated liquid: Atoms make transitions from the liquid phase to the solid phase, releasing energy and creating a new environment (the crystal) in which new quantum mechanical degrees of freedom (the phonons of a solid) can live; 
\item The transition from a paramagnet to a ferromagnet upon cooling below the Curie temperature: This process involves spontaneous symmetry breaking as one of the possible directions for the magnetization is chosen at random. Similarly to the measurement process, not a superposition of all possible outcomes but a specific outcome occurs. Such a phase transition requires a whole cascade of events until the macroscopic change is built up; 
\item The production of a flow of electrons by solar energy cells \cite{wurfel2005physics}: the activation of each electron due to light absorption is an event, and the subsequent conduction of electrons is a series of events, similar to what we described above in the context of the measurement process.
\item The damage that results to DNA molecules due to the impact of $\gamma$-ray photons, resulting in the possible development of cancer or changes in the phenotype of offspring;
developmental processes act as the amplifier to macro scales, see \cite{percival1991schrodinger};
\item The capture of a photon by a chlorophyll molecule, resulting in release of an electron that then results in a chain of biochemical reactions that eventually transform ADP to ATP and provide plants with the energy to grow; the transfer of energy in a Light Harvesting Complex is assisted by noise \cite{chin2010noise};
\item The quantum to classical transition of quantum perturbations in the inflationary era in the early universe, leading to classical perturbations on the Last Scattering surface rather then superpositions of such perturbations (\cite{Sudarsky2006},\cite{Sudarsky2015},\cite{Sudarsky2017},\cite{hollowood2017decoherence}), 
In order to describe this transition, a Lindblad equation is used in \cite{Burgess2015}.
\item The classical outcomes of nuclear reactions that occur during nucleosynthesis in the early universe \cite{Steigmann} (these processes do not result in a superposition of protons, deuterons, and helium states: they result in classical ratios of these nuclei).
\end{itemize}
 In each of these cases there is a macro context 
that guides what specific lower level events will take place \cite{EllisTopDown}.

\subsection{Issues arising}\label{sec:outcomes}

In this paper, we have presented a coherent and viable view that is in our opinion the most down-to-earth explanation of the measurement process.  
Due to the large variety of topics and questions that are relevant to our proposal, there arise many interesting issues that deserve further research. We conclude this paper by mentioning those issues that seem most important and relevant to us:
\begin{itemize}
\item \textbf{Entanglement on large scales and for many particles}:  The extraordinary achievement of macroscopic entanglement \cite{entangledZeil2007}\cite{entangleriedinger2018remote}\cite{entangled2018stabilized} has led many scientists to accept that quantum mechanics is universally valid. For instance, in \cite{gisin2017experimental},  single photons get absorbed in a crystal by a collective excitation of millions/billions of ions, and  the process is reversible as one can ``release'' the photon on demand. However, such situations are attained only by sufficiently isolating the system from interactions with the rest of the world, and in particular from interaction with heat baths. This requires low temperatures, or, in the case of long-distance entanglement experiments, time scales that are shorter than the characteristic time for interaction with a heat bath.
This in total contrast to the measurement process, where  interaction with the heat bath is the core of what is happening. By exploring the transition regime between these two limits one will certainly gain a better understanding of the crossover regime between the quantum and classical worlds. 
\item \textbf{Time scale over which unitary evolution acts} (eq.~(8) or eq.~(27)) before the coupling to the heat bath sets in. If this time interval is short enough, no integration is needed, but we have a series of collapses to the electron either still being in the ground state or being already in the excited state while the photon is still interacting with the system. If the time interval is not short enough, measuring deviations from Fermi's golden rule can be helpful in  exploring the time window between unitary evolution and collapse. 
\item \textbf{Necessary conditions for collapse}: Interaction with a heat bath as we defined it is in our view a sufficient condition for wave function collapse, or 'events'. However, there is no reason to expect that events cannot occur in systems that are not even locally in thermal equilibrium. We have mentioned criteria such as the uncontrollability of the microstate, feedback and nonlinearities, a quasi-continuum of states, and irreversibility due to emission of photon radiation as features of the heat bath that we consider important. Such features can also be realized in nonequilibrium systems.    
\item \textbf{Experimental confirmation}: As remarked in Section \ref{sec:intro_thispaper},  we have not added any \textit{ad hoc} terms into the Schr\"{o}dinger equation to attain our results; rather  
we have shown that effective collapse follows  from the established theory of open quantum systems when the local context is properly taken into account, provided that context includes heat baths as characterised above and that heat baths are defined the way we have done it. In that sense our proposal is already experimentally supported by the experiments  confirming the well-established  theory of open quantum systems. However more direct confirmation is desirable, and we are in discussions on how to test the proposals made here by specific experiments. 

\item \textbf{Quantum Field Theory}: We have confined ourselves to nonrelativistic physics. But the problem of wave function collapse arises also in QFT, as  all QFT calculations require preparation of an initial state and projection onto a final state. Creation and annihilation processes are in many situations understood to be ``real'' and not just a mathematical tool. One important question is: what is  the heat bath in particle production experiments? If one does not want to claim that the decision which particles are created is made in the detector surrounding the experiment, the relevant heat bath must be contained in the hot collision center. Similarly, early nucleosynthesis in cosmology \cite{Steigmann} requires a  suitable heat bath. This was provided by the cosmic plasma of baryons and radiation in equilibrium with each other at that time at a temperature of $\simeq 10^9$K \cite{dodelson2003} \cite{peter2013primordial}.

\end{itemize}

\noindent\textbf{Acknowledgements:} We thank Jean-Philippe Uzan for very helpful discussions at the start of the project. We thank Carlo Rovelli for useful comments on relational theories, 
and Richard Healey, Nicolas Gisin, Paul Davies, and Andrew Briggs for comments that led to clarification and extended discussion of certain points. GE thanks Rodolfo Gambini and Jorge Pullin for useful discussions on top-down effects and the quantum measurement problem that helped formulate his ideas.\\

GE thanks the National Research Foundation (South Africa) and the University of Cape Town Research Committee for financial support that assisted this work. This project began during a stay of BD at Perimeter Institute. Perimeter Institute is supported by the Government of Canada  through the Department of Innovation, Science and Economic Development and by the Province of Ontario through the Ministry of Research, Innovation and Science.
We furthermore acknowledge support by the German Research Foundation and the Open Access Publishing Fund of Technische Universit\"at Darmstadt.

\bibliographystyle{unsrt}


\end{document}